\documentclass[%twocolumn,onecolumn,12pt%
 reprint,
 amsmath,amssymb,
 aps,
prb,
]{revtex4-2}

\usepackage{tikz}
\usepackage{booktabs}
\usepackage{graphicx}% Include figure files
\usepackage{dcolumn}% Align table columns on decimal point
\usepackage{bm}% bold math
\begin{document}

\preprint{APS/123-QED}

\title{Spin Dynamics and Light-Induced Effects in EuZn$_2$P$_2$}% Force line breaks with \\

\author{M. Dutra$^1$}
\author{G. G. Vasques$^1$}
\author{P. C. Sabino$^1$}
\author{J. G. Dias$^1$}
\author{J. F. Oliveira$^2$}
\author{M. A. A. Heringer$^2$}
\author{M. Cabrera-Baez$^3$}
\author{E. Baggio Saitovitch$^2$}
\author{A. R. V. Benvenho$^1$}
\author{M. A. Avila$^1$}
 \author{J. Munevar$^{1,4}$},
  \email{Corresponding author: julian.munevar@correounivalle.edu.co;\\
  Present address: Universidad del Valle, Cali, A. A. 25360, Colombia.}

\affiliation{ $^1$CCNH, Universidade Federal do ABC (UFABC), Santo Andre, SP, 09210-580, Brazil}
 %weThis line break forced with \textbackslash\textbackslash
\affiliation{ $^2$Centro Brasileiro de Pesquisas Físicas, Rua Doutor Xavier Sigaud 150, Rio de Janeiro, RJ, 22290-180, Brazil}

\affiliation{$^3$Departamento de Física, Universidade Federal de Pernambuco, Recife, PE , 50670-901, Brazil}

\affiliation{$^4$Departamento de Física, Universidad del Valle, Cali, A. A. 25360, Colombia}

%\affiliation{
 %Third institution, the second for Charlie Author 
%\affiliation{%


\date{\today}% It is always \today, today,

\begin{abstract}

The magnetic spin dynamics and optical properties of EuZn$_2$P$_2$ are studied. 
Single crystals grown by the Sn-flux method crystallize in the $P\overline{3}m1$ (No.~164) space group and order antiferromagnetically at $T_N=23.5$~K. 
$^{151}$Eu Mössbauer spectroscopy confirms the presence of the Eu$^{2+}$ oxidation state only and the magnetic moment angle relative to the $c$-axis is $\theta=46(3)$\textdegree.
Temperature-dependent electron spin resonance (ESR) measurements reveal that spin-spin interactions predominantly govern the spin relaxation mechanisms, as evidenced by the linewidth behavior ($\Delta H$). 
Positive $g$-shifts ($\Delta g)$ for $H \parallel ab$ indicate the presence of local electron polarization. 
The ESR data support the formation of anisotropic magnetic polarons, which trap spin carriers and contribute to increased electrical resistance. 
Angular-dependent ESR spectra at room temperature display anisotropic behavior in both $\Delta g(\phi)$ and $\Delta H(\phi)$, with a dominant three-dimensional component $C_{3D}$, indicative of robust interlayer coupling and antiferromagnetic fluctuations. 
Under light illumination, a small broadening of $\Delta H$ is observed.
Furthermore, a photovoltaic effect is identified in EuZn$_2$P$_2$, with photodetector performance metrics suggesting promising capabilities for future optoelectronic devices.

%\begin{description}
%item[Structure]
%\end{description}
\end{abstract}

\maketitle

%\tableofcontents

\section{\label{sec:level1} Introduction}

Zintl phases, conceptualized through the Zintl-Klemm framework, have garnered significant attention as prospective thermoelectric materials that offer environmentally friendly pathways for energy conversion \cite{kauzlarich2023zintl}. 
This is in part due to the fact that many Zintl materials are known to have low environmental toxicity.
These materials are characterized by the transfer of electrons from electropositive to electronegative elements, stabilizing polyanionic frameworks.
One notable family within this class is represented by a subgroup of compounds with general formula RM$_2$X$_2$: those where R is an alkali-earth or rare-earth element, M is a $d^{5}$ or $d^{10}$ transition metal, and X is an element from groups 14 and 15 of the periodic table. 
They are stabilized by electron donation from the R element, thus satisfying the valence electron count required for structural integrity \cite{shuai2017recent, freer2022key}. 
Consequently, these compounds exhibit low electrical conductivity and are mainly semiconductors \cite{westbrook1995intermetallic}. 

Eu-based compounds with the CaAl$_2$Si$_2$-type structure, such as EuCd$_2$P$_2$ \cite{wang2021colossal}, EuCd$_2$As$_2$ \cite{ma2019spin}, EuZn$_2$As$_2$ \cite{luo2023colossal} exhibit colossal magnetoresistance (CMR) \cite{wang2021colossal, ma2019spin, du2022consecutive} and antiferromagnetic order at low temperatures. 
The observation of CMR in these materials is remarkable because it is usually observed in mixed-valence perovskites materials, which can exhibit ferromagnetism associated with double exchange or Jahn-Teller distortions \cite{littlewood2000transport}. 
Various mechanisms have been proposed to explain the CMR behavior, such as electronic band reconstruction \cite{zhang2023electronic}, spin-lattice coupling \cite{sunko2023spin}, and Berezinskii-Kosterlitz-Thouless (BKT) transitions \cite{heinrich2022topological}, among others.
These Eu-containing materials are also candidates for the realization of topological states.
For instance, EuZn$_2$As$_2$ has been proposed to exhibit a topological phase transition under hydrostatic pressure \cite{luo2023colossal}.
Moreover, EuCd$_2$As$_2$ has shown evidence of multiple topological transitions, as supported by pressure-dependent measurements and density functional theory (DFT) calculations \cite{du2022consecutive}. 
The interplay between magnetic and electronic degrees of freedom in these systems paves the way for novel quantum phenomena with potential applications in spintronics, quantum computing, and dark matter detection \cite{marsh2019proposal, ishiwata2024collective, schutte2021axion, sekine2021axion}.

In addition to their magnetic and electronic properties, Zintl phases can also be promising optoelectronic materials; tuning the band gap of these materials can make them viable candidates for photovoltaic applications.
Compounds such as SrCd$_2$X$_2$ (X = P, As) \cite{manzoor2023structural}, BaCd$_2$P$_2$ \cite{yuan2024discovery}, BaMg$_2$X$_2$ (X = P, As, Sb) \cite{souadi2024first}, Ba$_2$ZnP$_2$ \cite{khireddine2022elastic}, and YbZn$_2$X$_2$ (X = P, As, Sb, Bi) \cite{amin2024structural} have demonstrated optoelectronic properties that are potentially suitable for solar energy conversion.
However, to date, no photovoltaic effect has been reported in Eu-based CaAl$_2$Si$_2$-type compounds in single-crystalline form. 
The intersection of magnetism and light-induced effects remains largely unexplored in intermetallics, primarily due to the intrinsic challenges in coupling light with magnetic properties. 
Photons do not interact directly with magnetic fields as electrons do, making such interactions difficult to probe and observe.
Despite this, the unique potential of Eu-based CaAl$_2$Si$_2$-type compounds indicate that they may offer an ideal platform for studying the interplay between magnetism, optical excitation, and electronic behavior. 
When certain materials are exposed to illumination, electrons can undergo energy transitions.
%These changes affect the electron spin state, which can modify the magnetic properties of the material.
In such cases, photons absorbed by the electronic states can have a direct influence the magnetism. 
As the light is absorbed, it promotes the excitations of electrons into the conduction band or localized energy levels. 
In response, the electron density is redistributed and causes changes in the material magnetic properties \cite{kirilyuk2010ultrafast, kabychenkov1991magnetic}.
The photomagnetic effect can have relevant implications such as ultrafast control of magnetization \cite{kimel2005ultrafast}, photomagnetic sensing with high sensitivity and low detection limits \cite{dobrovolsky2005planar} and development of potential data storage devices with light-rewritable magnetic memory that does not require application of magnetic fields \cite{barrera2024light, kimel2019writing}.

In this work, we present experimental evidence of light-induced effects on the magnetism and spin dynamics of EuZn$_2$P$_2$, as well as the discovery of a photovoltaic effect.
These findings are supported by a combination of Mössbauer spectroscopy, ESR with and without light exposure, and room-temperature transport measurements under illumination.

\section{Methods}

High-purity single crystals of EuZn$_2$P$_2$ were synthesized by the Sn-flux method. 
High purity elements, Eu (99.9\%), Zn (99.999\%), P (99.999\%), and Sn (99.999\%) from Alfa-Aesar, were weighted in an atomic ratio of 1:2:2:40 and placed inside a quartz tube with quartz wool.
The evacuated and sealed ampoule was gradually heated to 500~ºC over 2~h, maintained for 1~h, then further heated to 1150~ºC over 4~h and held at this temperature for 10~h.  Controlled cooling was carried out down to 850~ºC at a rate of 2~ºC/h, after which the tubes were rapidly spun to separate the crystals from the flux.
The resulting single crystals were hexagonal in shape with approximately 1~mm x 1~mm x 0.5~mm. 
Powder X-ray diffraction (PXRD) was conducted at room temperature on crushed single crystals, using a Bruker D2 PHASER diffractometer with a LYNXEYE XE-T detector using Cu K$\alpha$ radiation with a wavelength of $\lambda_{Cu} = 1.5405$ \AA. 

Magnetic susceptibility measurements were carried out using a Quantum Design SQUID-MPMS3 system over the temperature range of 2-300~K.
Measurements were taken with the external magnetic field applied parallel and perpendicular to the crystallographic $c$-axis using applied fields of 1~kOe, 10~kOe and 30~kOe. 
The electrical resistance was measured in a Quantum Design Dynacool cryostat, using the Electrical Transport Option (ETO) in high-impedance mode. 

The ESR measurements were conducted using two setups: the temperature-dependent spectra were collected with a BRUKER - ELEXSYS 500 CW spectrometer equipped with a TE102 cavity operating at 9.4~GHz (X-band) using a continuous helium gas cryostat gas for temperatures from 18 to 250~K. 
The sample was aligned with $H \parallel ab$ inside a quartz straw before being inserted in the cryostat.
The magnetic field was swept from 1500 to 5500~Oe.
Angular-dependent ESR measurements were performed on an ESR5000 spectrometer using 2~mW microwave power and external field sweep from 500 to 6000~Oe. 
The modulation frequency was set to 10~kHz. 
Direct optical stimulation was performed by using an integrated light source, operated through ESRStudio for fastest response times. 

The $I-V$ curves were obtained using a Keithley 2602a source using different tensions with and without light.
The potentials used for the photovoltaic effect measurements were 0.5 and 15~V using a light source of 1000~lumens. The transient curves were obtained using a laser with a wavelength of 686~nm. 

\section{Results}

\subsection{Crystal Structure}

\begin{figure}[htpb]
   \centering
   \includegraphics[scale=.32]{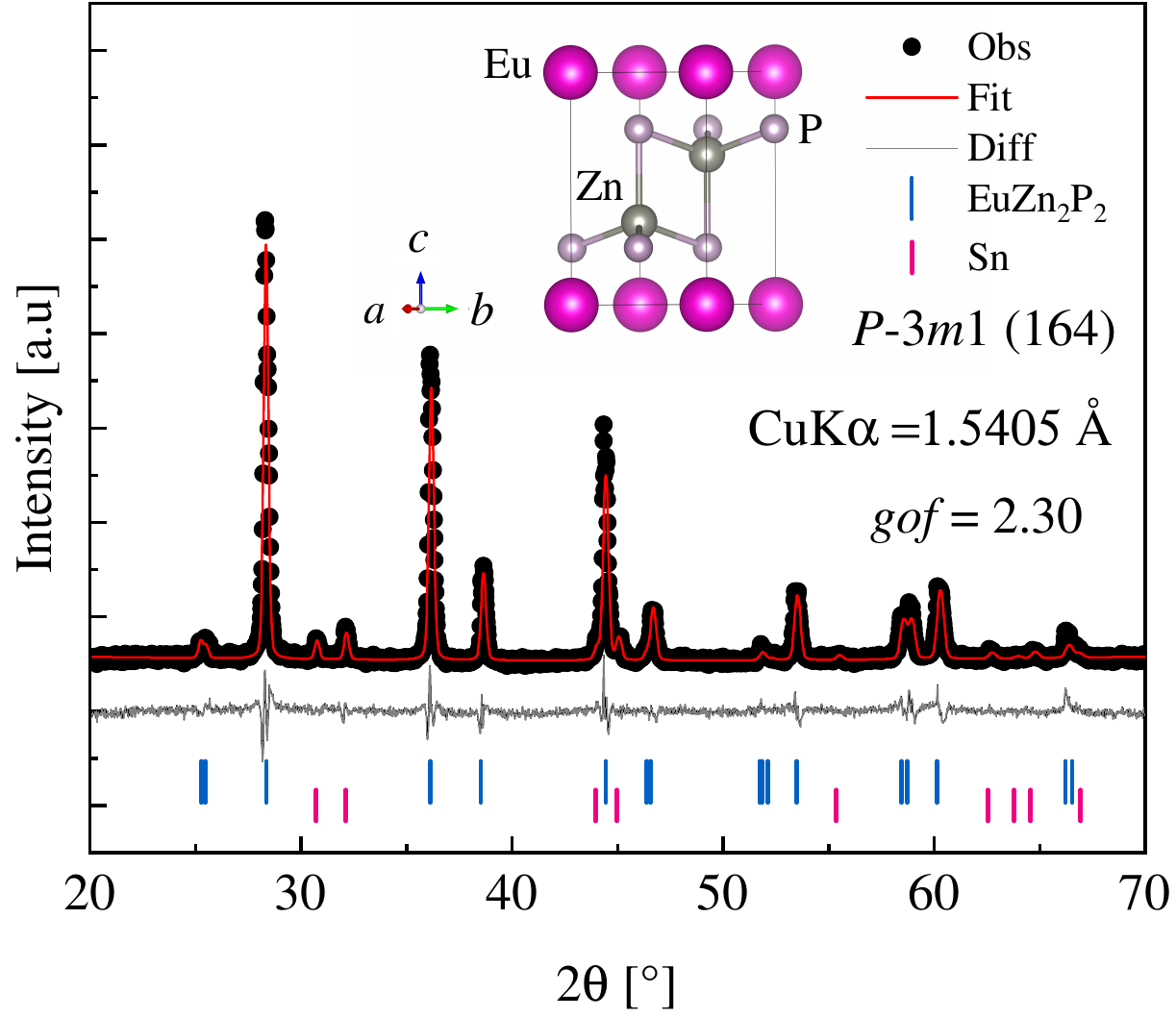}
   \caption{X-ray diffraction pattern of EuZn$_2$P$_2$. The experimental data is presented in black, the Rietveld refinement is shown in red, the difference between experiment and model is shown in gray, and the Bragg reflections corresponding to the Sn and EuZn$_2$P$_2$ are also shown as vertical lines. Inset are shown the EuZn$_2$P$_2$ unit cell. } 
   \label{fig:pxrd}
\end{figure}

The PXRD pattern and the Rietveld refinement obtained for the EuZn$_2$P$_2$ single crystals are shown in Fig.~\ref{fig:pxrd}. 
Rietveld refinement was performed using the GSAS-II software \cite{GSASII}, where a major phase was found corresponding to EuZn$_2$P$_2$, that crystallizes in a trigonal crystal lattice belonging to the space group $P\overline{3}m1$ (No.~164), plus a minor phase corresponding to a small amount of elemental Sn.
These Sn peaks are due to remaining flux on the EuZn$_2$P$_2$ crystal surface, which was mechanically removed prior to final crystal characterization.

\begin{figure*}[htpb]
   \centering
   \includegraphics[scale=.25]{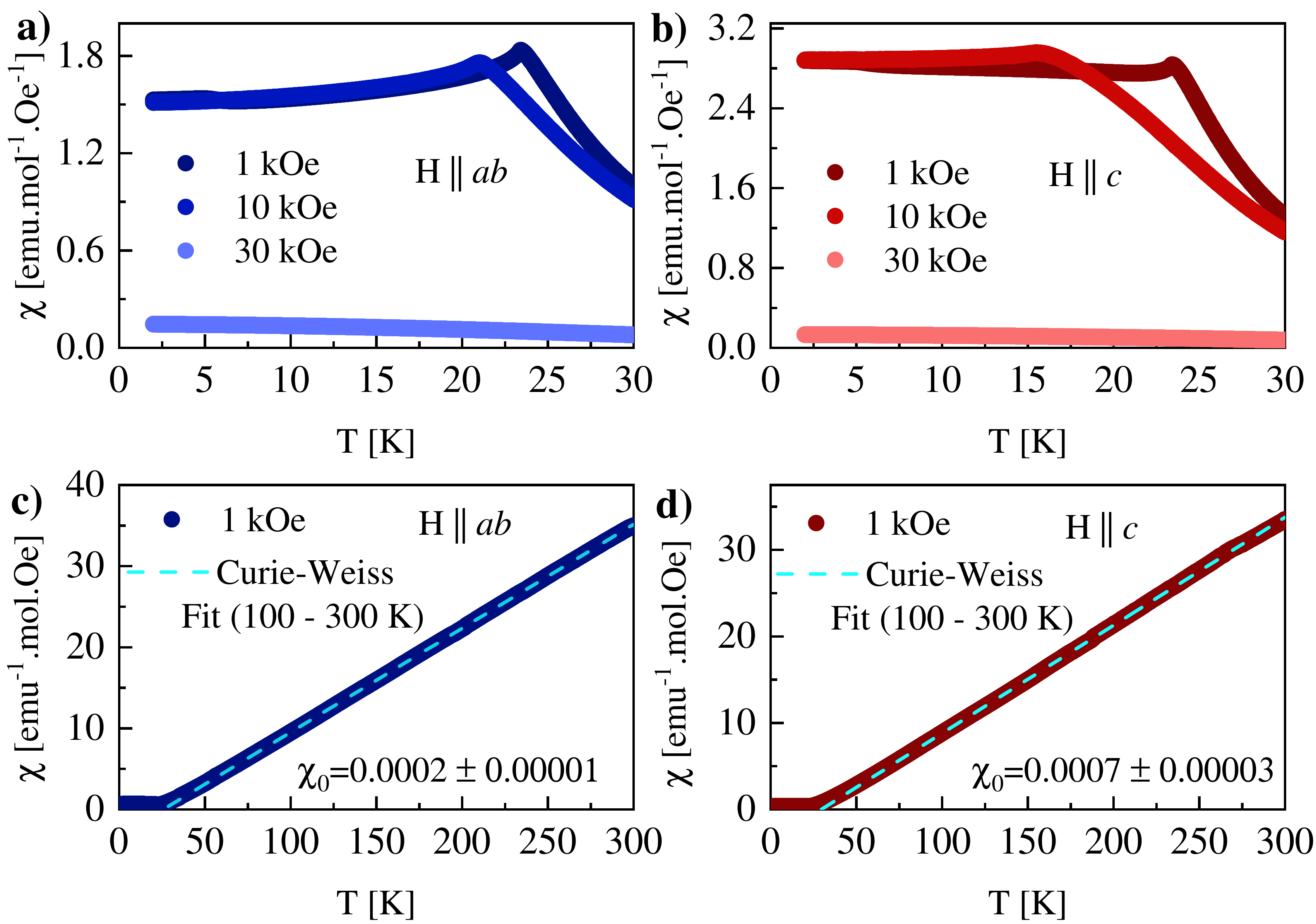}
  \caption{a) Magnetic susceptibility $\chi(T)$ for EuZn$_2$P$_2$ for $H \parallel ab$ and $H \parallel c$. It is possible to visualize the AFM ordering near $T_N=23.5$~K. As the applied field increase, the AFM order peak shifts to lower temperatures until its suppression. c) The Curie-Weiss analysis for $H \parallel ab$ and d) $H\parallel c$ on the $\chi^{-1}$ presenting linear behavior.} 
   \label{fig:mi}
\end{figure*}
   
The refined crystal structure (inset Fig.~\ref{fig:pxrd}), consists of Eu$^{2+}$ layers separated by [Zn$_2$P$_2$]$^{2-}$ clusters, consistent with the prototype structure CaAl$_2$Si$_2$.
The refined lattice parameters are $a = b = 4.0850(3)$~\AA, and $c = 7.0029(3)$~\AA, in good agreement with the literature \cite{berry2022dipolar}. 

\subsection{Magnetic Susceptibility}

Temperature-dependent magnetic susceptibility $\chi(T)$ was measured under various applied fields for both $H\parallel ab$ and $H\parallel c$ orientations (Fig.~\ref{fig:mi} (a-b)). 
An antiferromagnetic transition is observed at $T_N=23.5$~K, which is suppressed upon increasing the magnetic field.
The inverse susceptibility ($\chi^{-1}$) data fit to the Curie-Weiss law between 100 and 300~K (Fig.~\ref{fig:mi} (c-d)) yield  $\mu_{eff}^{H \parallel ab} = 7.85(9)$ $\mu_B$ for $H\parallel ab$ and $\mu_{eff}^{H \parallel c} = 8.11(7)$ for $H\parallel c$, with corresponding Curie-Weiss temperatures of $\theta_{CW}^{H \parallel ab} = 28.1(7)$~K  and  $\theta_{CW}^{H \parallel  c} = 27.9(8)$~K, respectively.
These positive values of $\theta_{CW}$ suggest the predominance of short-range FM interactions between Eu$^{2+}$ moments within each plane.

\subsection{$^{151}$Eu Mössbauer Spectroscopy}

\begin{figure}[htpb]
   \centering
   \includegraphics[scale=.45]{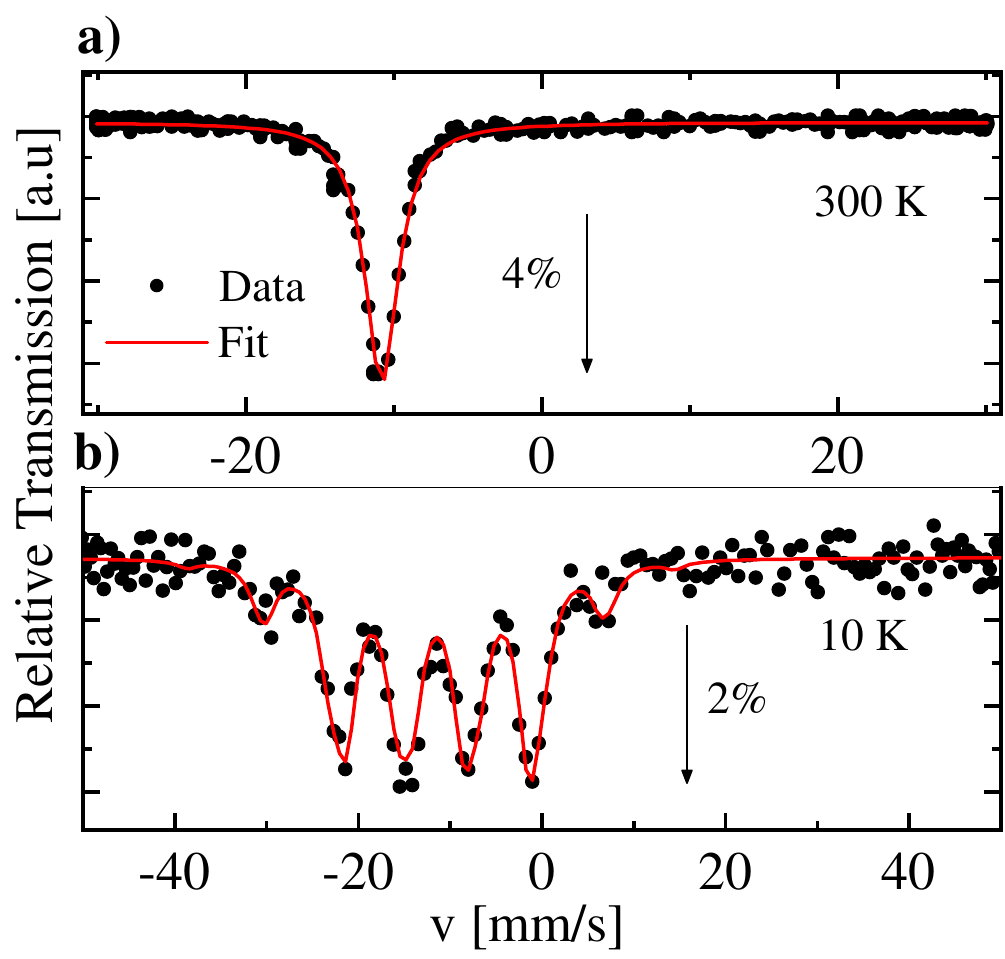}
   \caption{a) Mössbauer spectra for EuZn$_2$P$_2$ at 300~K and the fitting (red line). The room temperature spectrum shows unique resonance line with negative velocities indicating the Eu$^{2+}$ state. b) The 10~K spectrum present a series of resonance line related to the magnetic ordered state of Eu ion. } 
   \label{fig:m}
\end{figure}

$^{151}$Eu Mössbauer spectra obtained at 300~K and 10~K for EuZn$_2$P$_2$ crushed single crystals are shown in Fig.~\ref{fig:m}. 
Above $T_N$ a single broad resonance line is observed, whereas below $T_N$ the resonance line splits due to the onset of magnetic order.
The spectra were fitted using the Full Hamiltonian site analysis for the $^{151}$Eu nucleus, provided by the NORMOS software package. 

The fitted parameters are the isomer shift $\delta$, the quadrupole splitting $\Delta E_Q$, the linewidth $\Gamma$, the magnetic hyperfine field $B_{hf}$, the resonance area and the angle $\theta$ between $B_{hf}$ and the electric field gradient principal component $V_{zz}$, which has been taken parallel to the crystal $c$-axis.

At 300~K (Fig.~\ref{fig:m}a), a clear resonance line is obtained with a large negative isomer shift of $\delta=-11.1(9)$~mm/s, typical for Eu in the divalent state \cite{grandjean1989mossbauer}.
A small resonance asymmetry found in the spectrum at 300~K is attributed to the presence of a quadrupole splitting of $\Delta E_Q=2.5(4)$~mm/s.

Since no structural phase transitions are known in this compound and since the Eu$^{2+}$ has a half-filled $4f^7$ configuration, it is expected that the $4f$ electrons do not contribute to $V_{zz}$ and thus the contribution to $V_{zz}$ comes only from the crystal lattice.
Therefore, this value of $\Delta E_Q$ will be used as an input for the low-temperature spectra fit. 
No clear evidence for Eu$^{3+}$ or magnetic order are apparent in the 300~K spectrum.

The Mössbauer spectrum at 10~K (Fig.~\ref{fig:m}b) shows a resonance line splitting expected due to the onset of magnetic order of the Eu moments \cite{ahmida2019theoretical}.
The magnetically split spectrum fit was performed taking the $\Delta E_Q$ value obtained at 300~K, and the resulting hyperfine parameters are $\delta = - 11.4(1)$~mm/s, $B_{hf} = 25.2 \pm 0.1$~T, $\Gamma = 2.6(7)$~mm/s and $\theta=46(3)$\textdegree. 
Since $V_{zz}\parallel c$ is assumed, this means that the Eu moments are aligned at approximately 46\textdegree~from the $c$-axis, which is consistent with previous reports \cite{rybicki2024ambient}.

\subsection{Electron Spin Resonance}

\begin{figure*}[htpb]
   \centering
   \includegraphics[scale=.093]{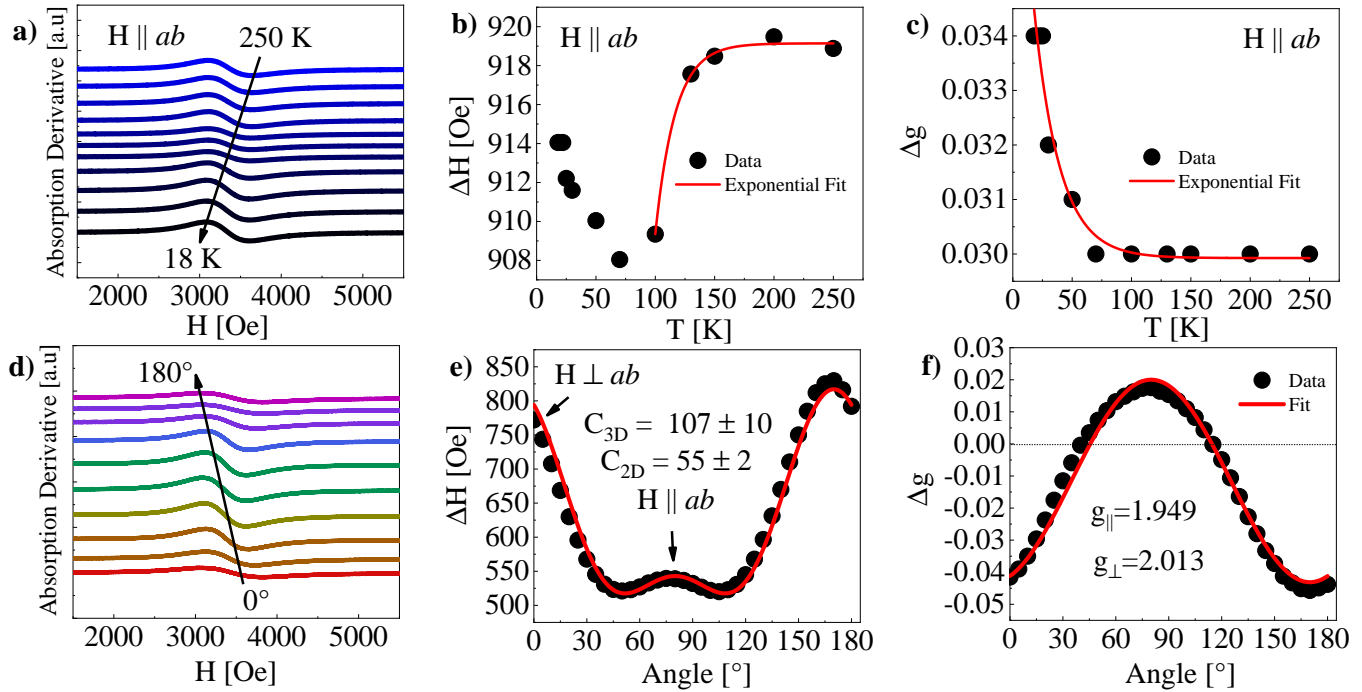}
  \caption{a) Temperature-dependent ESR spectra for $H \parallel ab$ b) $\Delta H(T)$ and c) $\Delta g(T)$ extracted using equation ~\ref{eq:L}. d) Angular-dependent ESR spectra at room-temperature e) $\Delta H(\phi)$ and f) $\Delta g(\phi)$.  } 
   \label{fig:esr}
\end{figure*}

ESR measurements (Fig.~\ref{fig:esr}a) were conducted on EuZn$_2$P$_2$ single crystals (18 - 250~K) with $H \parallel ab$. 
The ESR spectra reveal a single broad resonance line with small asymmetry and no hyperfine splitting, characteristic of localized magnetic moments in an exchanged-narrowed environment \cite{rosa2013magnetic}.
The spectra are well described by a Lorentzian admixture model combining absorption and dispersion components \cite{cabrera2015multiband}:
 
\begin{align}
     \frac{d [(1-\alpha)\chi"+\alpha \chi']}{dH}=
     \chi_0 H_0 \gamma_e^2T_2^2 \left[\frac{2(1-\alpha)x}{(1+x^2)^2}+\frac{\alpha(1-x^2)}{(1+x^2)^2}\right],
      \label{eq:L}
\end{align}
where
\begin{equation}
    x=(H_o - H)\gamma_e^2T_2,
\end{equation}

\noindent
$H_0$ and $H$ are, respectively, the resonance and the external field, $\gamma_e^2$ is the ratio of the electron gyromagnetic factor, $T_2$ is the spin-spin relaxation time, $\alpha$ is the admixture of absorption ($\alpha=0$) and dispersion ($\alpha=1$) and $\chi_0$ is the paramagnetic contribution from the static susceptibility. 
From these fits, the temperature evolution of the linewidth ($\Delta H$) and $g$-shift ($\Delta g$) was extracted as shown in Fig.~\ref{fig:esr} (b-c).
The $\Delta H(T)$ and $\Delta g(T)$ show exponential behavior, in good agreement with previous studies \cite{cook2025magnetic}.

We use the same approach to obtain the $\Delta H (\phi)$ and $\Delta g(\phi)$ from ESR spectra at room temperature for different angles between the crystal and the magnetic field (Fig.~\ref{fig:esr}d).
The extracted $\Delta H(\phi)$ and $\Delta g(\phi)$ (Fig.~\ref{fig:esr}e-f) display uniaxial behavior consistent with anisotropic magnetic interactions.
The model used to fit the $\Delta g(\phi)$ data considers two components of the $g$-factor ($g_{\parallel}$ and $g_{\perp}$) and is given by:

\begin{equation}
    g(\phi) = \sqrt{g_\parallel^2cos^2\phi +g_\perp^2sin^2\phi},
    \label{eq:gan}
\end{equation}
\noindent
where $\phi$ is the angle between the external applied field and the $c$-axis.
From the extracted components we note that $g_{\perp}>g_{\parallel}$, calculating the $\Delta g(\phi)$ yields both signs $\Delta g_{\perp}(\phi)>0$ and $\Delta g_{\parallel}(\phi)<0$.
This anisotropic Eu$^{2+}$ - Eu$^{2+}$ magnetic coupling is consistent with the scenario revealed in magnetic susceptibility data, where short-range FM interactions are in the $ab$-plane and AFM interactions are between layers.

Another interesting observation arises from the analysis of the angular dependence of the linewidth, $\Delta H(\phi)$.
The $\Delta H(\phi)$ data was extracted using models for 2D and 3D given by Eqs.~\ref{eq:3} and  \ref{eq:2}:

\begin{equation}
    \Delta H (\phi)_{3D} \propto C_{3D}(\cos^2(\phi)+1),
    \label{eq:3}
\end{equation}

\begin{equation}
    \Delta H (\phi)_{2D} \propto C_{2D}(3\cos^2(\phi)-1)^2,
    \label{eq:2}
\end{equation}

\noindent
where $\phi$ is the angle between the external magnetic field and the normal to the spin plane.
Spin fluctuations yield dominant contributions from the $C_{3D}$, indicating three-dimensional, short-wavelength spin fluctuations with significant interplane coupling and AFM fluctuations character, in good agreement with previous studies \cite{sichelschmidt2025electron}.

\subsection{Photoresponse}

\begin{figure*}[htpb]
   \centering
   \includegraphics[scale=.25]{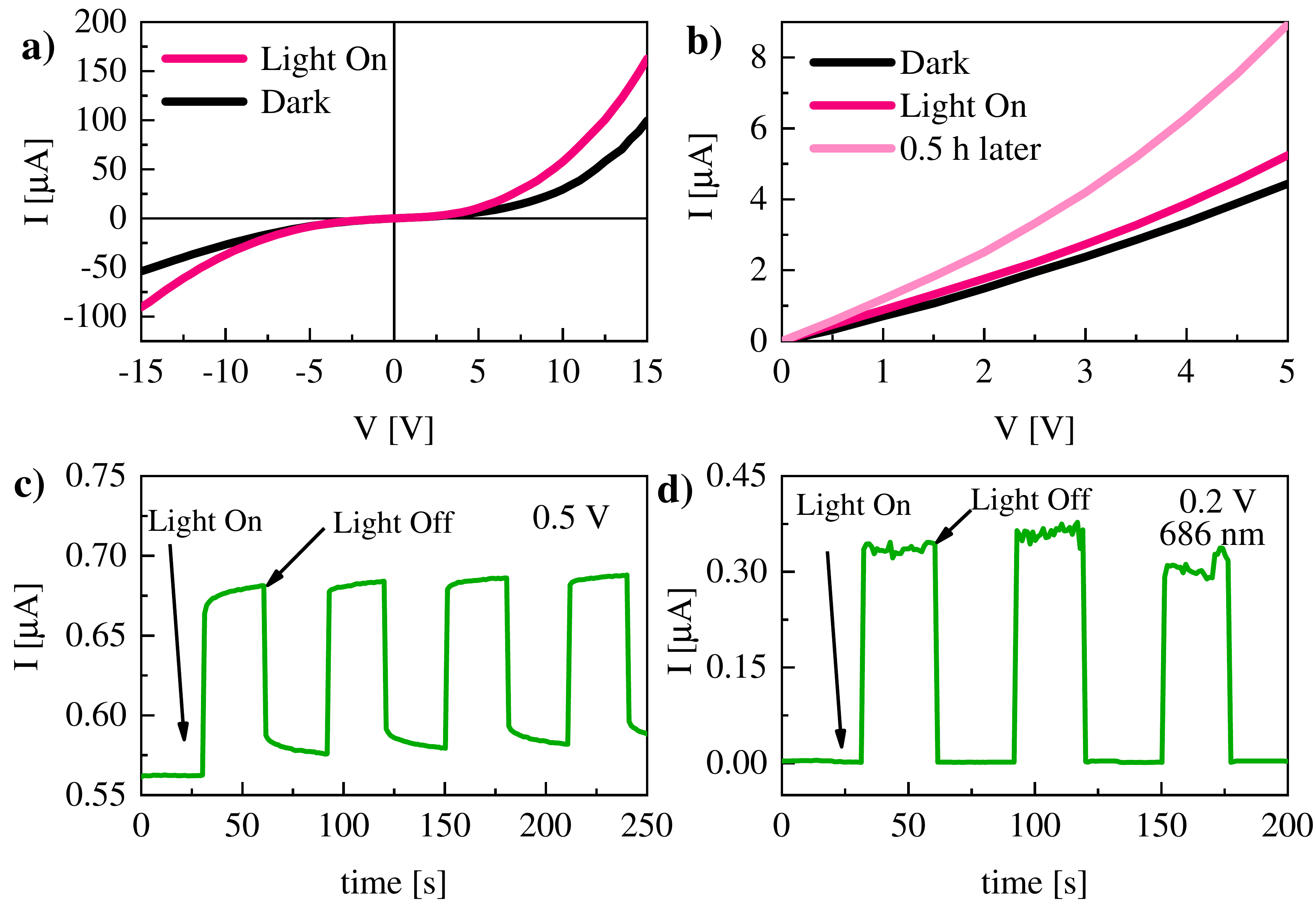}
  \caption{a) and b) are the $I-V$ curves without (dark) and with (purple) illumination. In d) we observe that the current persists to enhance after 0.5~h. c) Transient curves for 0.5~V and d) for 0~V for a laser with 696~nm wavelength with a  $t\sim 30$~s interval between the pulses is approximately.} 
   \label{fig:PV}
\end{figure*}

The photoresponse of EuZn$_2$P$_2$ was studied by analyzing the $I-V$ characteristics and transient response curves shown in Fig.~\ref{fig:PV}. 
The comparison of the $I-V$ curves under dark and illuminated conditions at 0.5~V (Fig.~\ref{fig:PV}a) reveals asymmetric and nonlinear behavior.
Such characteristics are present in certain photodiodes \cite{nikolic2011photodiode}.
The effect of light on the current is evident in Fig.~\ref{fig:PV}c, as the photogenerated current can be switched on and off by toggling the light source.

When a bias voltage of 0.5~V is applied, the dark current ($I_{Dark}$) starts at approximately 0.56~$\mu$A, which is relatively high even in the absence of illumination.
Upon exposure to light, the photocurrent increases significantly, reaching approximately 12~$\mu$A (Fig.~\ref{fig:PV}c).
To further investigate this effect, transient photocurrent measurements were carried out at zero bias using a laser with wavelength of approximately 686~nm (Fig.~\ref{fig:PV}d).
Under these conditions, the dark current remains practically negligible, while the photocurrent rises to around 0.35~$\mu$A upon laser illumination.

To investigate the possible effect of light radiation on EuZn$_2$P$_2$, we performed ESR measurements with 365, 462, 625~nm illumination and, from these spectra, we obtain the angle-dependent $\Delta H(\phi)$ and $\Delta g(\phi)$ under the effect of light.

\section{Discussion}

\subsection{Polaron formation}

The ESR-derived parameters (Fig.~\ref{fig:esr}) offer insight into the underlying spin dynamics in EuZn$_2$P$_2$. 
The $g$-shift $\Delta g$, defined as the difference between the experimental value $g_{exp}$ and the expected value for insulating Eu$^{2+}$ (1.993(2)) \cite{rosa2012electron},
can be expressed as:

\begin{equation}
    \Delta g =J_{fs}\eta(E_F),
    \label{eq:delta_g}
\end{equation}

\noindent
where $J_{fs}$ is the exchange interaction parameter between Eu$^{2+}$ ions and the $s$ spin carriers, and $\eta(E_F)$ denotes the density of states at the Fermi energy.

Since $J_{fs}$ and $\Delta g$ are related by Eq.~\ref{eq:delta_g}, this means that if $\Delta g < 0$ then an AFM polarization of spin carriers is expected, whereas $\Delta g > 0$ indicates FM polarization of spin carriers.
The persistent positive $\Delta g$ across the measured temperature range implies the presence of FM polarization of spin carriers surrounding the Eu$^{2+}$ ions \cite{abragam1970electron}. 
The observed enhancement of $\Delta g$ near $T_N$ reinforces the presence of short-range FM correlations, suggesting polaronic formation prior to long-range ordering \cite{cook2025magnetic}.

Magnetic polarons are quasiparticles formed by charge carriers which couple to the local magnetic moments via exchange interactions \cite{yakovlev2010magnetic}.
As a result, carriers become localized, and their wave functions are confined by this coupling. 
The main macroscopic consequence of this phenomenon is an increase in the resistivity \cite{littlewood2000transport}.

The evolution of $\Delta H$ further supports this scenario.
Between $70 < T < 200$~K the $\Delta H$ narrows.
However, as the system approaches $T_N$, $\Delta H$ broadens considerably, likely due to the onset of strong FM magnetic correlations. 
Such behavior is consistent with previous observations in systems where magnetic polarons emerge as a precursor to collective magnetic ordering \cite{rosa2012electron, souza2022microscopic}.
 
The existence of these localized states is related to an activated type of resistivity ($\rho\sim e^{\frac{\Delta}{k_BT}}$). 
As $T_N$ is reached, magnetic polarons begin to percolate; this percolation enables carrier transport, which in turn leads to a sharp drop in resistivity.
Based on our results, the formation of magnetic polarons in EuZn$_2$P$_2$ appears highly plausible.
First, we observe a minimum in $\Delta H$ around 50~K, which aligns closely with the resistivity peak reported in a previous study \cite{rybicki2024ambient}.
This minimum may be associated with the overlapping phase of magnetic polarons. 
Second, the evidence from $\Delta g$ also supports this scenario: the shift in $\Delta g$ is indicative of spin carrier polarization. 

\begin{figure}[htpb]
   \centering
   \includegraphics[scale=.33]{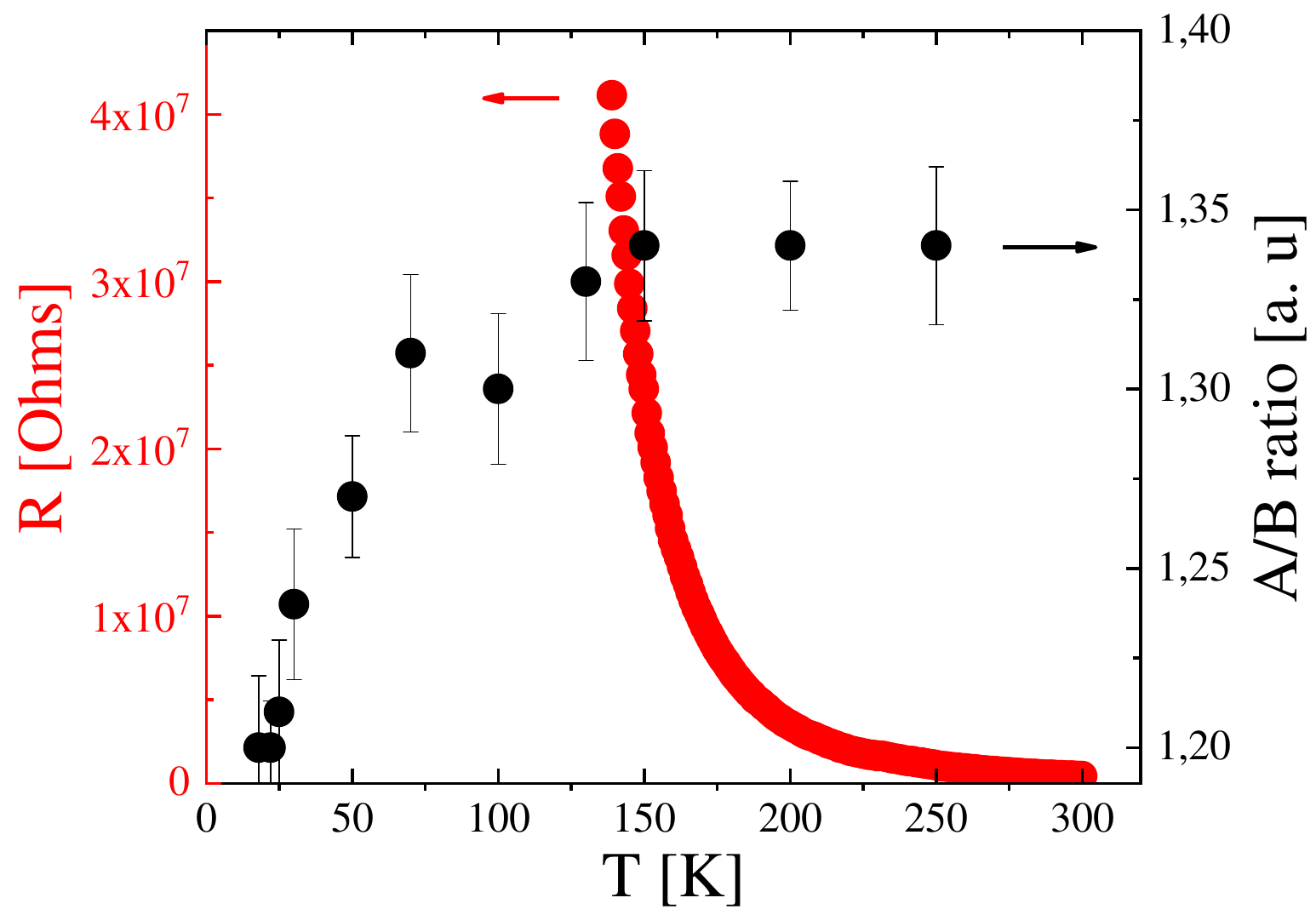}
   \caption{$A/B$ ratio (dark) and the electrical resistance plot (red). As the system in cooled, the ESR spectra tends to be more symmetric. As a consequence, we expect that the material should have a high insulating regime at low temperatures. } 
   \label{fig:ab}
\end{figure}

The $A/B$ ratio derived from the ESR line shape reveals a clear temperature dependence, becoming more symmetric at lower temperatures as shown in Fig.~\ref{fig:ab}.
This trend reflects a transition towards a more insulating state, in agreement with increased resistance, and is indicative of carrier localization \cite{cook2025magnetic}.
The correlation between ESR symmetry and resistance supports the interpretation of spin carriers being trapped within magnetic polarons, reducing the electrical conduction.  
These results collectively support a scenario wherein magnetic polarons in EuZn$_2$P$_2$ form anisotropically within a three-dimensional spin environment.
Although visualized as localized regions, the shape and extent of polarons may vary with magnetic anisotropy and spin dynamics.

With the possibility of magnetic polarons formation already established, we now turn our attention to light-dependent ESR measurements.

\subsection{Photoinduced effects}

\begin{figure}[htpb]
   \centering
   \includegraphics[scale=.22]{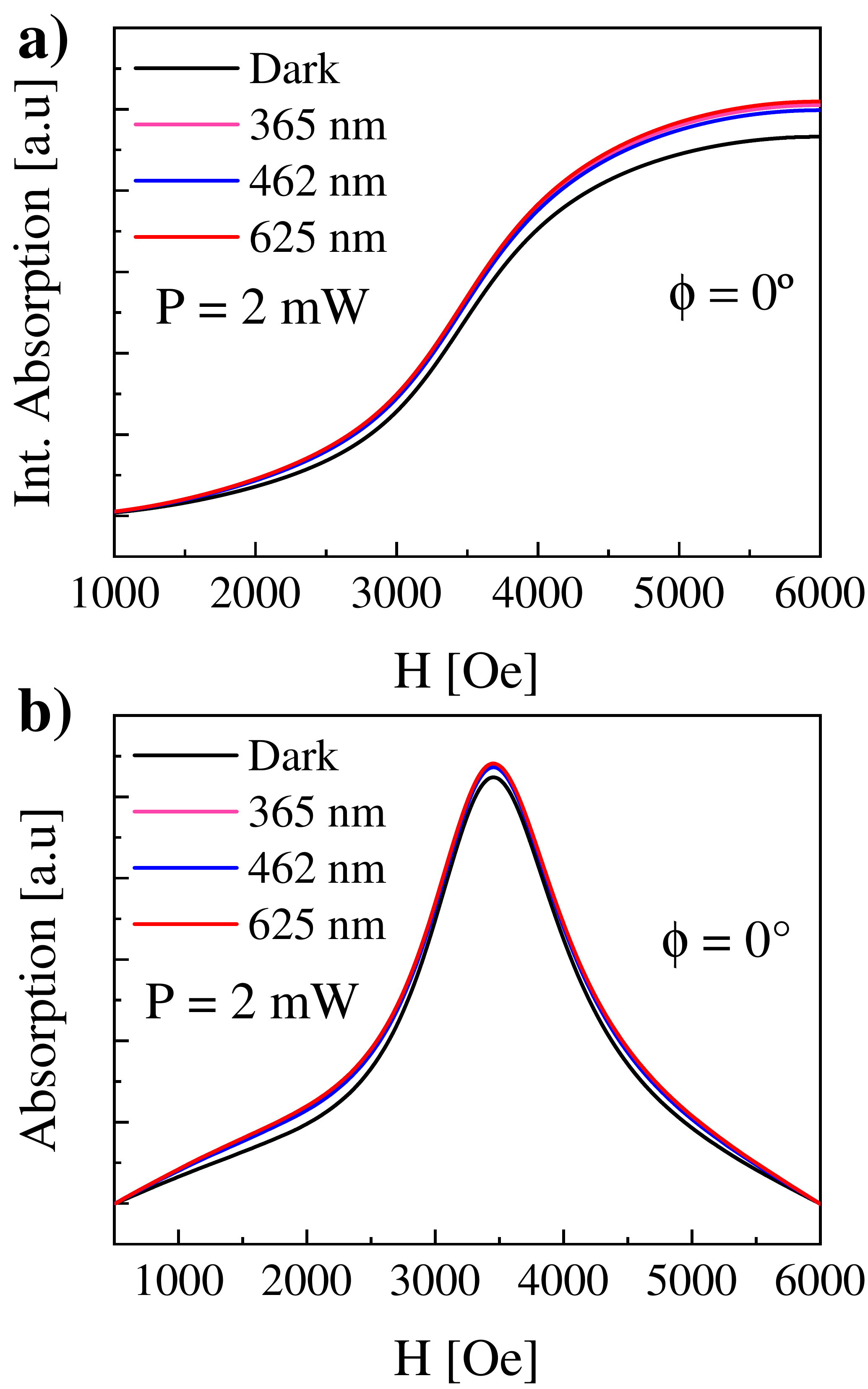}
   \caption{a) Integrated absorption and b) Absorption spectra  for the dark (black), 365 (pink), 462 (blue) and 625~nm (red) for a 2~mW power obtained from the room-temperature ESR spectra.} 
   \label{fig:int_EPR_light}
\end{figure}

The $I-V$ and transient response data (Fig.~\ref{fig:PV}) demonstrate a robust photovoltaic effect in EuZn$_2$P$_2$, with measurable increases in the current under visible light excitation.
The nonlinear, asymmetric behavior of the $I-V$ curves resembles that of semiconductor photodiodes, suggesting potential utility in optoelectronic applications.
To quantitatively assess the performance of EuZn$_2$P$_2$ as a photodetector, we must evaluate the key performance parameters:
responsivity ($R$), detectivity ($D$), and external quantum efficiency ($\eta_{ext}$) \cite{yang2022topological}:

\begin{equation}
    R=\frac{I_{photo}}{P},
\end{equation}
  
\begin{equation}
      D^{*} = R\sqrt{\frac{S}{2eI_{dark}}},
\end{equation}

\begin{equation}
  \eta_{external} =\frac{h\nu I_{photo}}{eP},
\end{equation}

\begin{figure*}[!htpb]
   \centering
   \includegraphics[scale=.18]{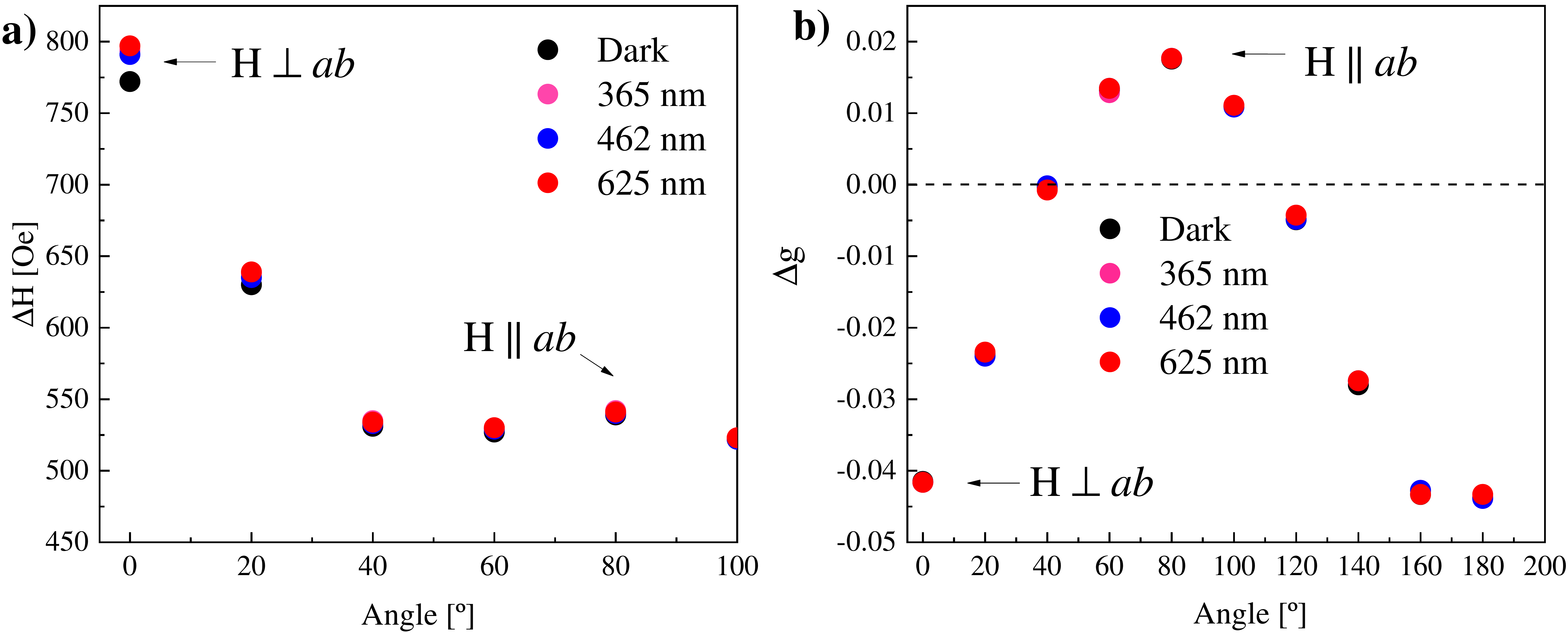}
   \caption{a) $\Delta H (\phi)$ and b) $\Delta g(\phi)$ without and with radiation. It's possible to observe small changes of the $\Delta H$ upon illumination but no significant in the $\Delta g (\phi)$. } 
   \label{fig:ligth}
\end{figure*}

\noindent
where $P$ is the power of the light source, $S$ is the effective surface area, and $I_{dark}$ and $I_{photo}$ are the dark current and the photogenerated electrical current, respectively.
For the 686~nm wavelength ($P\approx92.5$~mW), we obtain $R=3.78$~A/W, $D=1.39\times10^{11}$~Jones, and $\eta_{external}\approx 6$\%. These values are typical of a material with good performance as a photodetector, demonstrating potential for applications.

The ESR spectra under light irradiation with wavelengths of 365, 462 and 625~nm were analyzed to investigate the light-induced effects on the ESR spectra (Figs.~\ref{fig:int_EPR_light} and \ref{fig:ligth}).

A small broadening of $\Delta H$ is readily observed for the ESR spectra under the irradiation of visible light (Fig.~\ref{fig:ligth}a). 
As previously mentioned, such broadening may be indicative of more significant FM correlations. 
In this context, the light-induced ESR response provides two key insights:

\begin{enumerate}
    \item Light affects the magnetic behavior (by the $\Delta H$ light-induced broadening) of EuZn$_2$P$_2$; 
    \item Photoinduced electrons at the surface may play an active role in modulating the ESR signal.
\end{enumerate}

The $g$-factor, however, does not show any significant variation at least for the $g_{\perp}$ component (Fig.~\ref{fig:ligth}b).
This negligible change in $\Delta g$ is likely due to the limited penetration depth of the radiation. 
Since we use relatively low-energy photons, the light is not able to penetrate deeply into the bulk of the material.
As a result, the observed $g$-value, which primarily reflects the bulk contribution, is only weakly sensitive to changes in the surface region.
%Therefore, significant changes in $g$ are unlikely under these conditions. 

The ESR intensity, displayed in Fig.~\ref{fig:int_EPR_light}a, is directly proportional to the magnetic susceptibility and the population of resonant spins in the material. 
Thus, an increase in $I_{ESR}$ may indicate an increase in the spin population and/or an enhancement of the magnetic susceptibility. Interestingly, under the effect of light irradiation, we observe a clear increase in ESR intensity. 
This enhancement may be associated with photoinduced electron production at the surface of EuZn$_2$P$_2$ that contributes to the increased signal. 
This supports the idea that the radiation primarily affects only a few surface layers, but the effect remains readily detectable via ESR. 

Altogether, these light-induced effects in EuZn$_2$P$_2$ are manifestations of a relevant photomagnetic effect, which describes how the magnetic properties of a material (usually ferromagnetism) respond to light, typically leading to either enhancement or suppression of magnetization \cite{enz1969photomagnetic}.
However, this phenomenon is rarely observed in intermetallic compounds and is more commonly reported in organic and molecular magnetic systems \cite{garnier2016k, herrera2004reversible, ao1992electric, wu1994photo, o1987photo}.
Further experimental investigations will be useful to help clarify this effect in EuZn$_2$P$_2$.

\section{Conclusions}

We successfully grew high-quality EuZn$_2$P$_2$ single crystals using the Sn-flux method.
Mössbauer spectroscopy at 300~K shows a single resonance line at negative velocity, confirming the Eu$^{2+}$ oxidation state.
Analysis of the Mössbauer spectra reveals that the hyperfine field of Eu ions is canted by $\theta=46(3)$\textdegree ~ relative to the principal component of the crystal electric field (CEF).
ESR measurements demonstrate that the linewidth $\Delta H$ in EuZn$_2$P$_2$ is dominated by spin-spin interactions and exhibits $g$-shift $\Delta g > 0$.
These shifts are associated with the formation of internal fields caused by electron polarization around Eu$^{2+}$.
Such behavior, consistent with the $A/B$ ratio and electrical resistance plot, supports the presence of anisotropic magnetic polarons.
Angle-dependent analysis of $\Delta H(\phi)$ indicates that short-wavelength fluctuations are dominated by the $C_{3D}$ parameter, highlighting the three-dimensional nature of magnetic interactions in EuZn$_2$P$_2$ with strong interlayer coupling.
From the $\Delta g(\phi)$ analysis, we calculated $g{\parallel}$ and $g{\perp}$, finding that FM interactions are stronger along the $ab$-plane than in the $c$-axis. 
Specifically, $\Delta g_{\perp}$ is positive while $\Delta g_{\parallel}$ is negative.
ESR measurements under illumination show differences compared to dark conditions.
The linewidth $\Delta H$ broadens with light exposure, suggesting that light influences the magnetization of EuZn$_2$P$_2$.
Additionally, the ESR intensity increases under illumination, likely due to surface photoinduced electrons.
These observations point toward the possibility of a photomagnetic effect in EuZn$_2$P$_2$, although further experiments are necessary to confirm and detail its occurrence.

We also discovered a photovoltaic effect with promising performance parameters, characterized by responsivity, detectivity, and external quantum efficiency.
Taken together, magnetic polarons, photomagnetism, and photovoltaic behavior establish EuZn$_2$P$_2$ as a compelling platform to study the interplay between magnetism, optics, and electronics, with significant potential for applications in spintronics and quantum technologies.\\

\section{Acknowledgments}

The authors thanks P. G. Pagliuso and S. Sundar for fruitful discussions.
We also thank K. K. F. Barbosa and F. F. Ferreira for the PXRD measurements and the Multiuser Central Facilities at UFABC for the experimental support. 
We acknowledge the financial support of Brazilian funding agencies CAPES, CNPq (Contracts No. 140921/2022-2, No. 88887.837417/2023-00), FAPESP (No. 2017/20989-8, No. 2017/10581-1).
E. Baggio-Saitovitch, J.F. Oliveira, and M.A.V. Heringer thank Fundação Carlos Chagas Filho de Amparo a Pesquisa do Estado do Rio de Janeiro (FAPERJ) for Emeritus and PDN10 fellowships, and also for several grants including E-26/010.002990/2014 and E-26/210.496/2024.
M. Cabrera-Baez acknowledges the support of the INCT of Spintronics and Advanced Magnetic Nanostructures (INCT-SpinNanoMag), CNPq 406836/2022-1 and PROPESQI-UFPE.

%They turn out to be Eqs.~(\ref{appa}), (\ref{appb}), and (\ref{appc}).

\nocite{*}

\bibliography{apssamp}% Produces the bibliography via BibTeX.

\providecommand{\noopsort}[1]{}\providecommand{\singleletter}[1]{#1}%
\begin{thebibliography}{70}%
\makeatletter
\providecommand \@ifxundefined [1]{%
 \@ifx{#1\undefined}
}%
\providecommand \@ifnum [1]{%
 \ifnum #1\expandafter \@firstoftwo
 \else \expandafter \@secondoftwo
 \fi
}%
\providecommand \@ifx [1]{%
 \ifx #1\expandafter \@firstoftwo
 \else \expandafter \@secondoftwo
 \fi
}%
\providecommand \natexlab [1]{#1}%
\providecommand \enquote  [1]{``#1''}%
\providecommand \bibnamefont  [1]{#1}%
\providecommand \bibfnamefont [1]{#1}%
\providecommand \citenamefont [1]{#1}%
\providecommand \href@noop [0]{\@secondoftwo}%
\providecommand \href [0]{\begingroup \@sanitize@url \@href}%
\providecommand \@href[1]{\@@startlink{#1}\@@href}%
\providecommand \@@href[1]{\endgroup#1\@@endlink}%
\providecommand \@sanitize@url [0]{\catcode `\\12\catcode `\$12\catcode `\&12\catcode `\#12\catcode `\^12\catcode `\_12\catcode `\%12\relax}%
\providecommand \@@startlink[1]{}%
\providecommand \@@endlink[0]{}%
\providecommand \url  [0]{\begingroup\@sanitize@url \@url }%
\providecommand \@url [1]{\endgroup\@href {#1}{\urlprefix }}%
\providecommand \urlprefix  [0]{URL }%
\providecommand \Eprint [0]{\href }%
\providecommand \doibase [0]{https://doi.org/}%
\providecommand \selectlanguage [0]{\@gobble}%
\providecommand \bibinfo  [0]{\@secondoftwo}%
\providecommand \bibfield  [0]{\@secondoftwo}%
\providecommand \translation [1]{[#1]}%
\providecommand \BibitemOpen [0]{}%
\providecommand \bibitemStop [0]{}%
\providecommand \bibitemNoStop [0]{.\EOS\space}%
\providecommand \EOS [0]{\spacefactor3000\relax}%
\providecommand \BibitemShut  [1]{\csname bibitem#1\endcsname}%
\let\auto@bib@innerbib\@empty
%</preamble>
\bibitem [{\citenamefont {Kauzlarich}(2023)}]{kauzlarich2023zintl}%
  \BibitemOpen
  \bibfield  {author} {\bibinfo {author} {\bibfnamefont {S.~M.}\ \bibnamefont {Kauzlarich}},\ }\bibfield  {title} {\bibinfo {title} {Zintl phases: From curiosities to impactful materials},\ }\href@noop {} {\bibfield  {journal} {\bibinfo  {journal} {Chemistry of Materials}\ }\textbf {\bibinfo {volume} {35}},\ \bibinfo {pages} {7355} (\bibinfo {year} {2023})}\BibitemShut {NoStop}%
\bibitem [{\citenamefont {Shuai}\ \emph {et~al.}(2017)\citenamefont {Shuai}, \citenamefont {Mao}, \citenamefont {Song}, \citenamefont {Zhang}, \citenamefont {Chen},\ and\ \citenamefont {Ren}}]{shuai2017recent}%
  \BibitemOpen
  \bibfield  {author} {\bibinfo {author} {\bibfnamefont {J.}~\bibnamefont {Shuai}}, \bibinfo {author} {\bibfnamefont {J.}~\bibnamefont {Mao}}, \bibinfo {author} {\bibfnamefont {S.}~\bibnamefont {Song}}, \bibinfo {author} {\bibfnamefont {Q.}~\bibnamefont {Zhang}}, \bibinfo {author} {\bibfnamefont {G.}~\bibnamefont {Chen}},\ and\ \bibinfo {author} {\bibfnamefont {Z.}~\bibnamefont {Ren}},\ }\bibfield  {title} {\bibinfo {title} {Recent progress and future challenges on thermoelectric zintl materials},\ }\href@noop {} {\bibfield  {journal} {\bibinfo  {journal} {Materials Today Physics}\ }\textbf {\bibinfo {volume} {1}},\ \bibinfo {pages} {74} (\bibinfo {year} {2017})}\BibitemShut {NoStop}%
\bibitem [{\citenamefont {Freer}\ \emph {et~al.}(2022)\citenamefont {Freer}, \citenamefont {Ekren}, \citenamefont {Ghosh}, \citenamefont {Biswas}, \citenamefont {Qiu}, \citenamefont {Wan}, \citenamefont {Chen}, \citenamefont {Han}, \citenamefont {Fu}, \citenamefont {Zhu} \emph {et~al.}}]{freer2022key}%
  \BibitemOpen
  \bibfield  {author} {\bibinfo {author} {\bibfnamefont {R.}~\bibnamefont {Freer}}, \bibinfo {author} {\bibfnamefont {D.}~\bibnamefont {Ekren}}, \bibinfo {author} {\bibfnamefont {T.}~\bibnamefont {Ghosh}}, \bibinfo {author} {\bibfnamefont {K.}~\bibnamefont {Biswas}}, \bibinfo {author} {\bibfnamefont {P.}~\bibnamefont {Qiu}}, \bibinfo {author} {\bibfnamefont {S.}~\bibnamefont {Wan}}, \bibinfo {author} {\bibfnamefont {L.}~\bibnamefont {Chen}}, \bibinfo {author} {\bibfnamefont {S.}~\bibnamefont {Han}}, \bibinfo {author} {\bibfnamefont {C.}~\bibnamefont {Fu}}, \bibinfo {author} {\bibfnamefont {T.}~\bibnamefont {Zhu}}, \emph {et~al.},\ }\bibfield  {title} {\bibinfo {title} {Key properties of inorganic thermoelectric materials—tables (version 1)},\ }\href@noop {} {\bibfield  {journal} {\bibinfo  {journal} {Journal of Physics: Energy}\ }\textbf {\bibinfo {volume} {4}},\ \bibinfo {pages} {022002} (\bibinfo {year} {2022})}\BibitemShut {NoStop}%
\bibitem [{\citenamefont {Westbrook}\ and\ \citenamefont {Fleischer}(1995)}]{westbrook1995intermetallic}%
  \BibitemOpen
  \bibfield  {author} {\bibinfo {author} {\bibfnamefont {J.~H.}\ \bibnamefont {Westbrook}}\ and\ \bibinfo {author} {\bibfnamefont {R.~L.}\ \bibnamefont {Fleischer}},\ }\bibfield  {title} {\bibinfo {title} {Intermetallic compounds: principles and practice},\ }\href@noop {} {\bibfield  {journal} {\bibinfo  {journal} {(No Title)}\ } (\bibinfo {year} {1995})}\BibitemShut {NoStop}%
\bibitem [{\citenamefont {Wang}\ \emph {et~al.}(2021)\citenamefont {Wang}, \citenamefont {Rogers}, \citenamefont {Yao}, \citenamefont {Nichols}, \citenamefont {Atay}, \citenamefont {Xu}, \citenamefont {Franklin}, \citenamefont {Sochnikov}, \citenamefont {Ryan}, \citenamefont {Haskel} \emph {et~al.}}]{wang2021colossal}%
  \BibitemOpen
  \bibfield  {author} {\bibinfo {author} {\bibfnamefont {Z.-C.}\ \bibnamefont {Wang}}, \bibinfo {author} {\bibfnamefont {J.~D.}\ \bibnamefont {Rogers}}, \bibinfo {author} {\bibfnamefont {X.}~\bibnamefont {Yao}}, \bibinfo {author} {\bibfnamefont {R.}~\bibnamefont {Nichols}}, \bibinfo {author} {\bibfnamefont {K.}~\bibnamefont {Atay}}, \bibinfo {author} {\bibfnamefont {B.}~\bibnamefont {Xu}}, \bibinfo {author} {\bibfnamefont {J.}~\bibnamefont {Franklin}}, \bibinfo {author} {\bibfnamefont {I.}~\bibnamefont {Sochnikov}}, \bibinfo {author} {\bibfnamefont {P.~J.}\ \bibnamefont {Ryan}}, \bibinfo {author} {\bibfnamefont {D.}~\bibnamefont {Haskel}}, \emph {et~al.},\ }\bibfield  {title} {\bibinfo {title} {Colossal magnetoresistance without mixed valence in a layered phosphide crystal},\ }\href@noop {} {\bibfield  {journal} {\bibinfo  {journal} {Advanced Materials}\ }\textbf {\bibinfo {volume} {33}},\ \bibinfo {pages} {2005755} (\bibinfo {year} {2021})}\BibitemShut {NoStop}%
\bibitem [{\citenamefont {Ma}\ \emph {et~al.}(2019)\citenamefont {Ma}, \citenamefont {Nie}, \citenamefont {Yi}, \citenamefont {Jandke}, \citenamefont {Shang}, \citenamefont {Yao}, \citenamefont {Naamneh}, \citenamefont {Yan}, \citenamefont {Sun}, \citenamefont {Chikina} \emph {et~al.}}]{ma2019spin}%
  \BibitemOpen
  \bibfield  {author} {\bibinfo {author} {\bibfnamefont {J.-Z.}\ \bibnamefont {Ma}}, \bibinfo {author} {\bibfnamefont {S.}~\bibnamefont {Nie}}, \bibinfo {author} {\bibfnamefont {C.}~\bibnamefont {Yi}}, \bibinfo {author} {\bibfnamefont {J.}~\bibnamefont {Jandke}}, \bibinfo {author} {\bibfnamefont {T.}~\bibnamefont {Shang}}, \bibinfo {author} {\bibfnamefont {M.-Y.}\ \bibnamefont {Yao}}, \bibinfo {author} {\bibfnamefont {M.}~\bibnamefont {Naamneh}}, \bibinfo {author} {\bibfnamefont {L.}~\bibnamefont {Yan}}, \bibinfo {author} {\bibfnamefont {Y.}~\bibnamefont {Sun}}, \bibinfo {author} {\bibfnamefont {A.}~\bibnamefont {Chikina}}, \emph {et~al.},\ }\bibfield  {title} {\bibinfo {title} {Spin fluctuation induced weyl semimetal state in the paramagnetic phase of {EuCd$_2$As$_2$}},\ }\href@noop {} {\bibfield  {journal} {\bibinfo  {journal} {Science advances}\ }\textbf {\bibinfo {volume} {5}},\ \bibinfo {pages} {eaaw4718} (\bibinfo {year} {2019})}\BibitemShut {NoStop}%
\bibitem [{\citenamefont {Luo}\ \emph {et~al.}(2023)\citenamefont {Luo}, \citenamefont {Xu}, \citenamefont {Du}, \citenamefont {Yang}, \citenamefont {Chen}, \citenamefont {Cao}, \citenamefont {Song},\ and\ \citenamefont {Yuan}}]{luo2023colossal}%
  \BibitemOpen
  \bibfield  {author} {\bibinfo {author} {\bibfnamefont {S.}~\bibnamefont {Luo}}, \bibinfo {author} {\bibfnamefont {Y.}~\bibnamefont {Xu}}, \bibinfo {author} {\bibfnamefont {F.}~\bibnamefont {Du}}, \bibinfo {author} {\bibfnamefont {L.}~\bibnamefont {Yang}}, \bibinfo {author} {\bibfnamefont {Y.}~\bibnamefont {Chen}}, \bibinfo {author} {\bibfnamefont {C.}~\bibnamefont {Cao}}, \bibinfo {author} {\bibfnamefont {Y.}~\bibnamefont {Song}},\ and\ \bibinfo {author} {\bibfnamefont {H.}~\bibnamefont {Yuan}},\ }\bibfield  {title} {\bibinfo {title} {Colossal magnetoresistance and topological phase transition in {EuZn$_2$As$_2$}},\ }\href@noop {} {\bibfield  {journal} {\bibinfo  {journal} {Physical Review B}\ }\textbf {\bibinfo {volume} {108}},\ \bibinfo {pages} {205140} (\bibinfo {year} {2023})}\BibitemShut {NoStop}%
\bibitem [{\citenamefont {Du}\ \emph {et~al.}(2022)\citenamefont {Du}, \citenamefont {Yang}, \citenamefont {Nie}, \citenamefont {Wu}, \citenamefont {Li}, \citenamefont {Luo}, \citenamefont {Chen}, \citenamefont {Su}, \citenamefont {Smidman}, \citenamefont {Shi} \emph {et~al.}}]{du2022consecutive}%
  \BibitemOpen
  \bibfield  {author} {\bibinfo {author} {\bibfnamefont {F.}~\bibnamefont {Du}}, \bibinfo {author} {\bibfnamefont {L.}~\bibnamefont {Yang}}, \bibinfo {author} {\bibfnamefont {Z.}~\bibnamefont {Nie}}, \bibinfo {author} {\bibfnamefont {N.}~\bibnamefont {Wu}}, \bibinfo {author} {\bibfnamefont {Y.}~\bibnamefont {Li}}, \bibinfo {author} {\bibfnamefont {S.}~\bibnamefont {Luo}}, \bibinfo {author} {\bibfnamefont {Y.}~\bibnamefont {Chen}}, \bibinfo {author} {\bibfnamefont {D.}~\bibnamefont {Su}}, \bibinfo {author} {\bibfnamefont {M.}~\bibnamefont {Smidman}}, \bibinfo {author} {\bibfnamefont {Y.}~\bibnamefont {Shi}}, \emph {et~al.},\ }\bibfield  {title} {\bibinfo {title} {Consecutive topological phase transitions and colossal magnetoresistance in a magnetic topological semimetal},\ }\href@noop {} {\bibfield  {journal} {\bibinfo  {journal} {npj Quantum Materials}\ }\textbf {\bibinfo {volume} {7}},\ \bibinfo {pages} {65} (\bibinfo {year} {2022})}\BibitemShut {NoStop}%
\bibitem [{\citenamefont {Littlewood}(2000)}]{littlewood2000transport}%
  \BibitemOpen
  \bibfield  {author} {\bibinfo {author} {\bibfnamefont {P.}~\bibnamefont {Littlewood}},\ }\bibfield  {title} {\bibinfo {title} {Transport and magnetoresistance in low carrier density ferromagnets},\ }\href@noop {} {\bibfield  {journal} {\bibinfo  {journal} {Acta Physica Polonica A}\ }\textbf {\bibinfo {volume} {97}},\ \bibinfo {pages} {7} (\bibinfo {year} {2000})}\BibitemShut {NoStop}%
\bibitem [{\citenamefont {Zhang}\ \emph {et~al.}(2023)\citenamefont {Zhang}, \citenamefont {Du}, \citenamefont {Zheng}, \citenamefont {Luo}, \citenamefont {Wu}, \citenamefont {Zheng}, \citenamefont {Cui}, \citenamefont {Sun}, \citenamefont {Liu}, \citenamefont {Shen} \emph {et~al.}}]{zhang2023electronic}%
  \BibitemOpen
  \bibfield  {author} {\bibinfo {author} {\bibfnamefont {H.}~\bibnamefont {Zhang}}, \bibinfo {author} {\bibfnamefont {F.}~\bibnamefont {Du}}, \bibinfo {author} {\bibfnamefont {X.}~\bibnamefont {Zheng}}, \bibinfo {author} {\bibfnamefont {S.}~\bibnamefont {Luo}}, \bibinfo {author} {\bibfnamefont {Y.}~\bibnamefont {Wu}}, \bibinfo {author} {\bibfnamefont {H.}~\bibnamefont {Zheng}}, \bibinfo {author} {\bibfnamefont {S.}~\bibnamefont {Cui}}, \bibinfo {author} {\bibfnamefont {Z.}~\bibnamefont {Sun}}, \bibinfo {author} {\bibfnamefont {Z.}~\bibnamefont {Liu}}, \bibinfo {author} {\bibfnamefont {D.}~\bibnamefont {Shen}}, \emph {et~al.},\ }\bibfield  {title} {\bibinfo {title} {Electronic band reconstruction across the insulator-metal transition in colossally magnetoresistive {EuCd$_2$P$_2$}},\ }\href@noop {} {\bibfield  {journal} {\bibinfo  {journal} {Physical Review B}\ }\textbf {\bibinfo {volume} {108}},\ \bibinfo {pages} {L241115} (\bibinfo {year} {2023})}\BibitemShut {NoStop}%
\bibitem [{\citenamefont {Sunko}\ \emph {et~al.}(2023)\citenamefont {Sunko}, \citenamefont {Sun}, \citenamefont {Vranas}, \citenamefont {Homes}, \citenamefont {Lee}, \citenamefont {Donoway}, \citenamefont {Wang}, \citenamefont {Balguri}, \citenamefont {Mahendru}, \citenamefont {Ruiz} \emph {et~al.}}]{sunko2023spin}%
  \BibitemOpen
  \bibfield  {author} {\bibinfo {author} {\bibfnamefont {V.}~\bibnamefont {Sunko}}, \bibinfo {author} {\bibfnamefont {Y.}~\bibnamefont {Sun}}, \bibinfo {author} {\bibfnamefont {M.}~\bibnamefont {Vranas}}, \bibinfo {author} {\bibfnamefont {C.~C.}\ \bibnamefont {Homes}}, \bibinfo {author} {\bibfnamefont {C.}~\bibnamefont {Lee}}, \bibinfo {author} {\bibfnamefont {E.}~\bibnamefont {Donoway}}, \bibinfo {author} {\bibfnamefont {Z.-C.}\ \bibnamefont {Wang}}, \bibinfo {author} {\bibfnamefont {S.}~\bibnamefont {Balguri}}, \bibinfo {author} {\bibfnamefont {M.~B.}\ \bibnamefont {Mahendru}}, \bibinfo {author} {\bibfnamefont {A.}~\bibnamefont {Ruiz}}, \emph {et~al.},\ }\bibfield  {title} {\bibinfo {title} {Spin-carrier coupling induced ferromagnetism and giant resistivity peak in {EuCd$_2$P$_2$}},\ }\href@noop {} {\bibfield  {journal} {\bibinfo  {journal} {Physical Review B}\ }\textbf {\bibinfo {volume} {107}},\ \bibinfo {pages} {144404} (\bibinfo {year} {2023})}\BibitemShut {NoStop}%
\bibitem [{\citenamefont {Heinrich}\ \emph {et~al.}(2022)\citenamefont {Heinrich}, \citenamefont {Posske},\ and\ \citenamefont {Flebus}}]{heinrich2022topological}%
  \BibitemOpen
  \bibfield  {author} {\bibinfo {author} {\bibfnamefont {E.}~\bibnamefont {Heinrich}}, \bibinfo {author} {\bibfnamefont {T.}~\bibnamefont {Posske}},\ and\ \bibinfo {author} {\bibfnamefont {B.}~\bibnamefont {Flebus}},\ }\bibfield  {title} {\bibinfo {title} {Topological magnetic phase transition in {Eu}-based {A}-type antiferromagnets},\ }\href@noop {} {\bibfield  {journal} {\bibinfo  {journal} {Physical Review B}\ }\textbf {\bibinfo {volume} {106}},\ \bibinfo {pages} {214402} (\bibinfo {year} {2022})}\BibitemShut {NoStop}%
\bibitem [{\citenamefont {Marsh}\ \emph {et~al.}(2019)\citenamefont {Marsh}, \citenamefont {Fong}, \citenamefont {Lentz}, \citenamefont {{\v{S}}mejkal},\ and\ \citenamefont {Ali}}]{marsh2019proposal}%
  \BibitemOpen
  \bibfield  {author} {\bibinfo {author} {\bibfnamefont {D.~J.}\ \bibnamefont {Marsh}}, \bibinfo {author} {\bibfnamefont {K.~C.}\ \bibnamefont {Fong}}, \bibinfo {author} {\bibfnamefont {E.~W.}\ \bibnamefont {Lentz}}, \bibinfo {author} {\bibfnamefont {L.}~\bibnamefont {{\v{S}}mejkal}},\ and\ \bibinfo {author} {\bibfnamefont {M.~N.}\ \bibnamefont {Ali}},\ }\bibfield  {title} {\bibinfo {title} {Proposal to detect dark matter using axionic topological antiferromagnets},\ }\href@noop {} {\bibfield  {journal} {\bibinfo  {journal} {Physical Review Letters}\ }\textbf {\bibinfo {volume} {123}},\ \bibinfo {pages} {121601} (\bibinfo {year} {2019})}\BibitemShut {NoStop}%
\bibitem [{\citenamefont {Ishiwata}\ and\ \citenamefont {Nomura}(2024)}]{ishiwata2024collective}%
  \BibitemOpen
  \bibfield  {author} {\bibinfo {author} {\bibfnamefont {K.}~\bibnamefont {Ishiwata}}\ and\ \bibinfo {author} {\bibfnamefont {K.}~\bibnamefont {Nomura}},\ }\bibfield  {title} {\bibinfo {title} {Collective excitations in magnetic topological insulators and axion dark matter search},\ }\href@noop {} {\bibfield  {journal} {\bibinfo  {journal} {arXiv preprint arXiv:2406.09705}\ } (\bibinfo {year} {2024})}\BibitemShut {NoStop}%
\bibitem [{\citenamefont {Sch{\"u}tte-Engel}\ \emph {et~al.}(2021)\citenamefont {Sch{\"u}tte-Engel}, \citenamefont {Marsh}, \citenamefont {Millar}, \citenamefont {Sekine}, \citenamefont {Chadha-Day}, \citenamefont {Hoof}, \citenamefont {Ali}, \citenamefont {Fong}, \citenamefont {Hardy},\ and\ \citenamefont {{\v{S}}mejkal}}]{schutte2021axion}%
  \BibitemOpen
  \bibfield  {author} {\bibinfo {author} {\bibfnamefont {J.}~\bibnamefont {Sch{\"u}tte-Engel}}, \bibinfo {author} {\bibfnamefont {D.~J.}\ \bibnamefont {Marsh}}, \bibinfo {author} {\bibfnamefont {A.~J.}\ \bibnamefont {Millar}}, \bibinfo {author} {\bibfnamefont {A.}~\bibnamefont {Sekine}}, \bibinfo {author} {\bibfnamefont {F.}~\bibnamefont {Chadha-Day}}, \bibinfo {author} {\bibfnamefont {S.}~\bibnamefont {Hoof}}, \bibinfo {author} {\bibfnamefont {M.~N.}\ \bibnamefont {Ali}}, \bibinfo {author} {\bibfnamefont {K.~C.}\ \bibnamefont {Fong}}, \bibinfo {author} {\bibfnamefont {E.}~\bibnamefont {Hardy}},\ and\ \bibinfo {author} {\bibfnamefont {L.}~\bibnamefont {{\v{S}}mejkal}},\ }\bibfield  {title} {\bibinfo {title} {Axion quasiparticles for axion dark matter detection},\ }\href@noop {} {\bibfield  {journal} {\bibinfo  {journal} {Journal of Cosmology and Astroparticle Physics}\ }\textbf {\bibinfo {volume} {2021}}\bibinfo  {number} { (08)},\ \bibinfo {pages} {066}}\BibitemShut {NoStop}%
\bibitem [{\citenamefont {Sekine}\ and\ \citenamefont {Nomura}(2021)}]{sekine2021axion}%
  \BibitemOpen
\bibfield  {number} {  }\bibfield  {author} {\bibinfo {author} {\bibfnamefont {A.}~\bibnamefont {Sekine}}\ and\ \bibinfo {author} {\bibfnamefont {K.}~\bibnamefont {Nomura}},\ }\bibfield  {title} {\bibinfo {title} {Axion electrodynamics in topological materials},\ }\href@noop {} {\bibfield  {journal} {\bibinfo  {journal} {Journal of Applied Physics}\ }\textbf {\bibinfo {volume} {129}} (\bibinfo {year} {2021})}\BibitemShut {NoStop}%
\bibitem [{\citenamefont {Manzoor}\ \emph {et~al.}(2023)\citenamefont {Manzoor}, \citenamefont {Behera}, \citenamefont {Sharma}, \citenamefont {Iqbal}, \citenamefont {Mukherjee}, \citenamefont {Khenata}, \citenamefont {Bin-Omran}, \citenamefont {Alshahrani}, \citenamefont {El~Shiekh},\ and\ \citenamefont {Ouahrani}}]{manzoor2023structural}%
  \BibitemOpen
  \bibfield  {author} {\bibinfo {author} {\bibfnamefont {M.}~\bibnamefont {Manzoor}}, \bibinfo {author} {\bibfnamefont {D.}~\bibnamefont {Behera}}, \bibinfo {author} {\bibfnamefont {R.}~\bibnamefont {Sharma}}, \bibinfo {author} {\bibfnamefont {M.~W.}\ \bibnamefont {Iqbal}}, \bibinfo {author} {\bibfnamefont {S.}~\bibnamefont {Mukherjee}}, \bibinfo {author} {\bibfnamefont {R.}~\bibnamefont {Khenata}}, \bibinfo {author} {\bibfnamefont {S.}~\bibnamefont {Bin-Omran}}, \bibinfo {author} {\bibfnamefont {T.}~\bibnamefont {Alshahrani}}, \bibinfo {author} {\bibfnamefont {E.}~\bibnamefont {El~Shiekh}},\ and\ \bibinfo {author} {\bibfnamefont {T.}~\bibnamefont {Ouahrani}},\ }\bibfield  {title} {\bibinfo {title} {Structural, electronic, optical, and thermoelectric studies on zintl {SrCd$_2$Pn$_2$ (Pn= P/As)} compounds for solar cell applications: A first principle approach},\ }\href@noop {} {\bibfield  {journal} {\bibinfo  {journal} {Journal of Solid State Chemistry}\ }\textbf {\bibinfo {volume} {326}},\ \bibinfo {pages}
  {124188} (\bibinfo {year} {2023})}\BibitemShut {NoStop}%
\bibitem [{\citenamefont {Yuan}\ \emph {et~al.}(2024)\citenamefont {Yuan}, \citenamefont {Dahliah}, \citenamefont {Hasan}, \citenamefont {Kassa}, \citenamefont {Pike}, \citenamefont {Quadir}, \citenamefont {Claes}, \citenamefont {Chandler}, \citenamefont {Xiong}, \citenamefont {Kyveryga} \emph {et~al.}}]{yuan2024discovery}%
  \BibitemOpen
  \bibfield  {author} {\bibinfo {author} {\bibfnamefont {Z.}~\bibnamefont {Yuan}}, \bibinfo {author} {\bibfnamefont {D.}~\bibnamefont {Dahliah}}, \bibinfo {author} {\bibfnamefont {M.~R.}\ \bibnamefont {Hasan}}, \bibinfo {author} {\bibfnamefont {G.}~\bibnamefont {Kassa}}, \bibinfo {author} {\bibfnamefont {A.}~\bibnamefont {Pike}}, \bibinfo {author} {\bibfnamefont {S.}~\bibnamefont {Quadir}}, \bibinfo {author} {\bibfnamefont {R.}~\bibnamefont {Claes}}, \bibinfo {author} {\bibfnamefont {C.}~\bibnamefont {Chandler}}, \bibinfo {author} {\bibfnamefont {Y.}~\bibnamefont {Xiong}}, \bibinfo {author} {\bibfnamefont {V.}~\bibnamefont {Kyveryga}}, \emph {et~al.},\ }\bibfield  {title} {\bibinfo {title} {Discovery of the zintl-phosphide {BaCd$_2$P$_2$} as a long carrier lifetime and stable solar absorber},\ }\href@noop {} {\bibfield  {journal} {\bibinfo  {journal} {Joule}\ }\textbf {\bibinfo {volume} {8}},\ \bibinfo {pages} {1412} (\bibinfo {year} {2024})}\BibitemShut {NoStop}%
\bibitem [{\citenamefont {Souadi}(2024)}]{souadi2024first}%
  \BibitemOpen
  \bibfield  {author} {\bibinfo {author} {\bibfnamefont {G.}~\bibnamefont {Souadi}},\ }\bibfield  {title} {\bibinfo {title} {First principles investigations of optoelectronic and thermoelectric properties of novel {BaMg$_2$X$_2$ (X= P, As, Sb)} alloys for renewable energy applications},\ }\href@noop {} {\bibfield  {journal} {\bibinfo  {journal} {Inorganic Chemistry Communications}\ ,\ \bibinfo {pages} {112768}} (\bibinfo {year} {2024})}\BibitemShut {NoStop}%
\bibitem [{\citenamefont {Khireddine}\ \emph {et~al.}(2022)\citenamefont {Khireddine}, \citenamefont {Bouhemadou}, \citenamefont {Maabed}, \citenamefont {Bin-Omran}, \citenamefont {Khenata},\ and\ \citenamefont {Al-Douri}}]{khireddine2022elastic}%
  \BibitemOpen
  \bibfield  {author} {\bibinfo {author} {\bibfnamefont {A.}~\bibnamefont {Khireddine}}, \bibinfo {author} {\bibfnamefont {A.}~\bibnamefont {Bouhemadou}}, \bibinfo {author} {\bibfnamefont {S.}~\bibnamefont {Maabed}}, \bibinfo {author} {\bibfnamefont {S.}~\bibnamefont {Bin-Omran}}, \bibinfo {author} {\bibfnamefont {R.}~\bibnamefont {Khenata}},\ and\ \bibinfo {author} {\bibfnamefont {Y.}~\bibnamefont {Al-Douri}},\ }\bibfield  {title} {\bibinfo {title} {Elastic, electronic, optical and thermoelectric properties of the novel zintl-phase {Ba$_2$ZnP$_2$}},\ }\href@noop {} {\bibfield  {journal} {\bibinfo  {journal} {Solid State Sciences}\ }\textbf {\bibinfo {volume} {128}},\ \bibinfo {pages} {106893} (\bibinfo {year} {2022})}\BibitemShut {NoStop}%
\bibitem [{\citenamefont {Amin}\ \emph {et~al.}(2024)\citenamefont {Amin}, \citenamefont {Reshak}, \citenamefont {Zada}, \citenamefont {ur~Rahman}, \citenamefont {Ali}, \citenamefont {Khan}, \citenamefont {Laref}, \citenamefont {Ramli} \emph {et~al.}}]{amin2024structural}%
  \BibitemOpen
  \bibfield  {author} {\bibinfo {author} {\bibfnamefont {T.}~\bibnamefont {Amin}}, \bibinfo {author} {\bibfnamefont {A.~H.}\ \bibnamefont {Reshak}}, \bibinfo {author} {\bibfnamefont {Z.}~\bibnamefont {Zada}}, \bibinfo {author} {\bibfnamefont {I.}~\bibnamefont {ur~Rahman}}, \bibinfo {author} {\bibfnamefont {D.}~\bibnamefont {Ali}}, \bibinfo {author} {\bibfnamefont {A.~M.}\ \bibnamefont {Khan}}, \bibinfo {author} {\bibfnamefont {A.}~\bibnamefont {Laref}}, \bibinfo {author} {\bibfnamefont {M.~M.}\ \bibnamefont {Ramli}}, \emph {et~al.},\ }\bibfield  {title} {\bibinfo {title} {Structural, opto-electronic and transport properties of zintl compound {YbZn$_2$Y$_2$ (Y= P, As, Sb, Bi)}},\ }\href@noop {} {\bibfield  {journal} {\bibinfo  {journal} {Engineered Science}\ }\textbf {\bibinfo {volume} {31}},\ \bibinfo {pages} {1258} (\bibinfo {year} {2024})}\BibitemShut {NoStop}%
\bibitem [{\citenamefont {Kirilyuk}\ \emph {et~al.}(2010)\citenamefont {Kirilyuk}, \citenamefont {Kimel},\ and\ \citenamefont {Rasing}}]{kirilyuk2010ultrafast}%
  \BibitemOpen
  \bibfield  {author} {\bibinfo {author} {\bibfnamefont {A.}~\bibnamefont {Kirilyuk}}, \bibinfo {author} {\bibfnamefont {A.~V.}\ \bibnamefont {Kimel}},\ and\ \bibinfo {author} {\bibfnamefont {T.}~\bibnamefont {Rasing}},\ }\bibfield  {title} {\bibinfo {title} {Ultrafast optical manipulation of magnetic order},\ }\href@noop {} {\bibfield  {journal} {\bibinfo  {journal} {Reviews of Modern Physics}\ }\textbf {\bibinfo {volume} {82}},\ \bibinfo {pages} {2731} (\bibinfo {year} {2010})}\BibitemShut {NoStop}%
\bibitem [{\citenamefont {Kabychenkov}(1991)}]{kabychenkov1991magnetic}%
  \BibitemOpen
  \bibfield  {author} {\bibinfo {author} {\bibfnamefont {A.}~\bibnamefont {Kabychenkov}},\ }\bibfield  {title} {\bibinfo {title} {Magnetic phase transitions in a light wave},\ }\href@noop {} {\bibfield  {journal} {\bibinfo  {journal} {Sov. Phys. JETP}\ }\textbf {\bibinfo {volume} {73}},\ \bibinfo {pages} {672} (\bibinfo {year} {1991})}\BibitemShut {NoStop}%
\bibitem [{\citenamefont {Kimel}\ \emph {et~al.}(2005)\citenamefont {Kimel}, \citenamefont {Kirilyuk}, \citenamefont {Usachev}, \citenamefont {Pisarev}, \citenamefont {Balbashov},\ and\ \citenamefont {Rasing}}]{kimel2005ultrafast}%
  \BibitemOpen
  \bibfield  {author} {\bibinfo {author} {\bibfnamefont {A.}~\bibnamefont {Kimel}}, \bibinfo {author} {\bibfnamefont {A.}~\bibnamefont {Kirilyuk}}, \bibinfo {author} {\bibfnamefont {P.}~\bibnamefont {Usachev}}, \bibinfo {author} {\bibfnamefont {R.}~\bibnamefont {Pisarev}}, \bibinfo {author} {\bibfnamefont {A.}~\bibnamefont {Balbashov}},\ and\ \bibinfo {author} {\bibfnamefont {T.}~\bibnamefont {Rasing}},\ }\bibfield  {title} {\bibinfo {title} {Ultrafast non-thermal control of magnetization by instantaneous photomagnetic pulses},\ }\href@noop {} {\bibfield  {journal} {\bibinfo  {journal} {Nature}\ }\textbf {\bibinfo {volume} {435}},\ \bibinfo {pages} {655} (\bibinfo {year} {2005})}\BibitemShut {NoStop}%
\bibitem [{\citenamefont {Dobrovolsky}\ and\ \citenamefont {Rossokhaty}(2005)}]{dobrovolsky2005planar}%
  \BibitemOpen
  \bibfield  {author} {\bibinfo {author} {\bibfnamefont {V.}~\bibnamefont {Dobrovolsky}}\ and\ \bibinfo {author} {\bibfnamefont {V.}~\bibnamefont {Rossokhaty}},\ }\bibfield  {title} {\bibinfo {title} {Planar photomagnetic effect soi sensors for various applications with low detection limit},\ }in\ \href@noop {} {\emph {\bibinfo {booktitle} {Science and Technology of Semiconductor-On-Insulator Structures and Devices Operating in a Harsh Environment: Proceedings of the NATO Advanced Research Workshop on Science and Technology of Semiconductor-On-Insulator Structures and Devices Operating in a Harsh Environment Kiev, Ukraine 26--30 April 2004}}}\ (\bibinfo {organization} {Springer},\ \bibinfo {year} {2005})\ pp.\ \bibinfo {pages} {303--308}\BibitemShut {NoStop}%
\bibitem [{\citenamefont {Barrera}\ \emph {et~al.}(2024)\citenamefont {Barrera}, \citenamefont {Martella}, \citenamefont {Celegato}, \citenamefont {Fuochi}, \citenamefont {Co{\"\i}sson}, \citenamefont {Parmeggiani}, \citenamefont {Wiersma},\ and\ \citenamefont {Tiberto}}]{barrera2024light}%
  \BibitemOpen
  \bibfield  {author} {\bibinfo {author} {\bibfnamefont {G.}~\bibnamefont {Barrera}}, \bibinfo {author} {\bibfnamefont {D.}~\bibnamefont {Martella}}, \bibinfo {author} {\bibfnamefont {F.}~\bibnamefont {Celegato}}, \bibinfo {author} {\bibfnamefont {N.}~\bibnamefont {Fuochi}}, \bibinfo {author} {\bibfnamefont {M.}~\bibnamefont {Co{\"\i}sson}}, \bibinfo {author} {\bibfnamefont {C.}~\bibnamefont {Parmeggiani}}, \bibinfo {author} {\bibfnamefont {D.~S.}\ \bibnamefont {Wiersma}},\ and\ \bibinfo {author} {\bibfnamefont {P.}~\bibnamefont {Tiberto}},\ }\bibfield  {title} {\bibinfo {title} {Light-controlled magnetic properties: An energy-efficient opto-mechanical control over magnetic films by liquid crystalline networks},\ }\href@noop {} {\bibfield  {journal} {\bibinfo  {journal} {Advanced Science}\ }\textbf {\bibinfo {volume} {11}},\ \bibinfo {pages} {2408273} (\bibinfo {year} {2024})}\BibitemShut {NoStop}%
\bibitem [{\citenamefont {Kimel}\ and\ \citenamefont {Li}(2019)}]{kimel2019writing}%
  \BibitemOpen
  \bibfield  {author} {\bibinfo {author} {\bibfnamefont {A.~V.}\ \bibnamefont {Kimel}}\ and\ \bibinfo {author} {\bibfnamefont {M.}~\bibnamefont {Li}},\ }\bibfield  {title} {\bibinfo {title} {Writing magnetic memory with ultrashort light pulses},\ }\href@noop {} {\bibfield  {journal} {\bibinfo  {journal} {Nature Reviews Materials}\ }\textbf {\bibinfo {volume} {4}},\ \bibinfo {pages} {189} (\bibinfo {year} {2019})}\BibitemShut {NoStop}%
\bibitem [{\citenamefont {Toby}\ and\ \citenamefont {Von~Dreele}(2013)}]{GSASII}%
  \BibitemOpen
  \bibfield  {author} {\bibinfo {author} {\bibfnamefont {B.~H.}\ \bibnamefont {Toby}}\ and\ \bibinfo {author} {\bibfnamefont {R.~B.}\ \bibnamefont {Von~Dreele}},\ }\bibfield  {title} {\bibinfo {title} {Gsas-ii: the genesis of a modern open-source all purpose crystallography software package},\ }\href {https://doi.org/https://doi.org/10.1107/S0021889813003531} {\bibfield  {journal} {\bibinfo  {journal} {Journal of Applied Crystallography}\ }\textbf {\bibinfo {volume} {46}},\ \bibinfo {pages} {544} (\bibinfo {year} {2013})}\BibitemShut {NoStop}%
\bibitem [{\citenamefont {Berry}\ \emph {et~al.}(2022)\citenamefont {Berry}, \citenamefont {Stewart}, \citenamefont {Redemann}, \citenamefont {Lygouras}, \citenamefont {Varnava}, \citenamefont {Vanderbilt},\ and\ \citenamefont {McQueen}}]{berry2022dipolar}%
  \BibitemOpen
  \bibfield  {author} {\bibinfo {author} {\bibfnamefont {T.}~\bibnamefont {Berry}}, \bibinfo {author} {\bibfnamefont {V.~J.}\ \bibnamefont {Stewart}}, \bibinfo {author} {\bibfnamefont {B.~W.}\ \bibnamefont {Redemann}}, \bibinfo {author} {\bibfnamefont {C.}~\bibnamefont {Lygouras}}, \bibinfo {author} {\bibfnamefont {N.}~\bibnamefont {Varnava}}, \bibinfo {author} {\bibfnamefont {D.}~\bibnamefont {Vanderbilt}},\ and\ \bibinfo {author} {\bibfnamefont {T.~M.}\ \bibnamefont {McQueen}},\ }\bibfield  {title} {\bibinfo {title} {Dipolar magnetic interactions and {A}-type antiferromagnetic order in the zintl phase insulator {EuZn$_2$P$_2$}},\ }\href@noop {} {\bibfield  {journal} {\bibinfo  {journal} {arXiv preprint arXiv:2203.12739}\ } (\bibinfo {year} {2022})}\BibitemShut {NoStop}%
\bibitem [{\citenamefont {Grandjean}\ and\ \citenamefont {Long}(1989)}]{grandjean1989mossbauer}%
  \BibitemOpen
  \bibfield  {author} {\bibinfo {author} {\bibfnamefont {F.}~\bibnamefont {Grandjean}}\ and\ \bibinfo {author} {\bibfnamefont {G.~J.}\ \bibnamefont {Long}},\ }\bibfield  {title} {\bibinfo {title} {M{\"o}ssbauer spectroscopy of europium-containing compounds},\ }in\ \href@noop {} {\emph {\bibinfo {booktitle} {M{\"o}ssbauer Spectroscopy Applied to Inorganic Chemistry}}}\ (\bibinfo  {publisher} {Springer},\ \bibinfo {year} {1989})\ pp.\ \bibinfo {pages} {513--597}\BibitemShut {NoStop}%
\bibitem [{\citenamefont {Ahmida}\ \emph {et~al.}(2019)\citenamefont {Ahmida} \emph {et~al.}}]{ahmida2019theoretical}%
  \BibitemOpen
  \bibfield  {author} {\bibinfo {author} {\bibfnamefont {M.~A.}\ \bibnamefont {Ahmida}} \emph {et~al.},\ }\bibfield  {title} {\bibinfo {title} {Theoretical review of m{\"o}ssbauer effect, hyperfine interactions parameters and the valence fluctuations in eu systems},\ }\href@noop {} {\bibfield  {journal} {\bibinfo  {journal} {Journal of Applied Mathematics and Physics}\ }\textbf {\bibinfo {volume} {7}},\ \bibinfo {pages} {254} (\bibinfo {year} {2019})}\BibitemShut {NoStop}%
\bibitem [{\citenamefont {Rybicki}\ \emph {et~al.}(2024)\citenamefont {Rybicki}, \citenamefont {Kom{\c{e}}dera}, \citenamefont {Przewo{\'z}nik}, \citenamefont {Gondek}, \citenamefont {Kapusta}, \citenamefont {Podg{\'o}rska}, \citenamefont {Tabi{\'s}}, \citenamefont {{\.Z}ukrowski}, \citenamefont {Tran}, \citenamefont {Babij} \emph {et~al.}}]{rybicki2024ambient}%
  \BibitemOpen
  \bibfield  {author} {\bibinfo {author} {\bibfnamefont {D.}~\bibnamefont {Rybicki}}, \bibinfo {author} {\bibfnamefont {K.}~\bibnamefont {Kom{\c{e}}dera}}, \bibinfo {author} {\bibfnamefont {J.}~\bibnamefont {Przewo{\'z}nik}}, \bibinfo {author} {\bibfnamefont {{\L}.}~\bibnamefont {Gondek}}, \bibinfo {author} {\bibfnamefont {C.}~\bibnamefont {Kapusta}}, \bibinfo {author} {\bibfnamefont {K.}~\bibnamefont {Podg{\'o}rska}}, \bibinfo {author} {\bibfnamefont {W.}~\bibnamefont {Tabi{\'s}}}, \bibinfo {author} {\bibfnamefont {J.}~\bibnamefont {{\.Z}ukrowski}}, \bibinfo {author} {\bibfnamefont {L.~M.}\ \bibnamefont {Tran}}, \bibinfo {author} {\bibfnamefont {M.}~\bibnamefont {Babij}}, \emph {et~al.},\ }\bibfield  {title} {\bibinfo {title} {Ambient-and high-pressure studies of structural, electronic, and magnetic properties of single-crystal {EuZn$_2$P$_2$}},\ }\href@noop {} {\bibfield  {journal} {\bibinfo  {journal} {Physical Review B}\ }\textbf {\bibinfo {volume} {110}},\ \bibinfo {pages} {014421} (\bibinfo {year}
  {2024})}\BibitemShut {NoStop}%
\bibitem [{\citenamefont {Rosa}\ \emph {et~al.}(2013)\citenamefont {Rosa}, \citenamefont {Iwamoto}, \citenamefont {Holanda}, \citenamefont {Ribeiro}, \citenamefont {Pagliuso}, \citenamefont {Rettori},\ and\ \citenamefont {Avila}}]{rosa2013magnetic}%
  \BibitemOpen
  \bibfield  {author} {\bibinfo {author} {\bibfnamefont {P.~F.~S.}\ \bibnamefont {Rosa}}, \bibinfo {author} {\bibfnamefont {W.}~\bibnamefont {Iwamoto}}, \bibinfo {author} {\bibfnamefont {L.}~\bibnamefont {Holanda}}, \bibinfo {author} {\bibfnamefont {R.}~\bibnamefont {Ribeiro}}, \bibinfo {author} {\bibfnamefont {P.}~\bibnamefont {Pagliuso}}, \bibinfo {author} {\bibfnamefont {C.}~\bibnamefont {Rettori}},\ and\ \bibinfo {author} {\bibfnamefont {M.}~\bibnamefont {Avila}},\ }\bibfield  {title} {\bibinfo {title} {Magnetic polaron effect in {Sr$_{8-x}$Eu$_{x}$Ga$_{16}$Ge$_{30}$} clathrates probed by electron spin resonance},\ }\href@noop {} {\bibfield  {journal} {\bibinfo  {journal} {Physical Review B—Condensed Matter and Materials Physics}\ }\textbf {\bibinfo {volume} {87}},\ \bibinfo {pages} {224414} (\bibinfo {year} {2013})}\BibitemShut {NoStop}%
\bibitem [{\citenamefont {Cabrera-Baez}\ \emph {et~al.}(2015)\citenamefont {Cabrera-Baez}, \citenamefont {Naranjo-Uribe}, \citenamefont {Osorio-Guill{\'e}n}, \citenamefont {Rettori},\ and\ \citenamefont {Avila}}]{cabrera2015multiband}%
  \BibitemOpen
  \bibfield  {author} {\bibinfo {author} {\bibfnamefont {M.}~\bibnamefont {Cabrera-Baez}}, \bibinfo {author} {\bibfnamefont {A.}~\bibnamefont {Naranjo-Uribe}}, \bibinfo {author} {\bibfnamefont {J.}~\bibnamefont {Osorio-Guill{\'e}n}}, \bibinfo {author} {\bibfnamefont {C.}~\bibnamefont {Rettori}},\ and\ \bibinfo {author} {\bibfnamefont {M.}~\bibnamefont {Avila}},\ }\bibfield  {title} {\bibinfo {title} {Multiband electronic characterization of the complex intermetallic cage system {Y$_{1-x}$Gd$_x$Co$_2$Zn$_20$}},\ }\href@noop {} {\bibfield  {journal} {\bibinfo  {journal} {Physical Review B}\ }\textbf {\bibinfo {volume} {92}},\ \bibinfo {pages} {214414} (\bibinfo {year} {2015})}\BibitemShut {NoStop}%
\bibitem [{\citenamefont {Cook}\ \emph {et~al.}(2025)\citenamefont {Cook}, \citenamefont {Peterson}, \citenamefont {Kengle}, \citenamefont {Kennedy}, \citenamefont {Sheeran}, \citenamefont {Girod}, \citenamefont {Freitas}, \citenamefont {Greer}, \citenamefont {Abbamonte}, \citenamefont {Pagliuso} \emph {et~al.}}]{cook2025magnetic}%
  \BibitemOpen
  \bibfield  {author} {\bibinfo {author} {\bibfnamefont {M.~S.}\ \bibnamefont {Cook}}, \bibinfo {author} {\bibfnamefont {E.~A.}\ \bibnamefont {Peterson}}, \bibinfo {author} {\bibfnamefont {C.~S.}\ \bibnamefont {Kengle}}, \bibinfo {author} {\bibfnamefont {E.}~\bibnamefont {Kennedy}}, \bibinfo {author} {\bibfnamefont {J.}~\bibnamefont {Sheeran}}, \bibinfo {author} {\bibfnamefont {C.}~\bibnamefont {Girod}}, \bibinfo {author} {\bibfnamefont {G.}~\bibnamefont {Freitas}}, \bibinfo {author} {\bibfnamefont {S.~M.}\ \bibnamefont {Greer}}, \bibinfo {author} {\bibfnamefont {P.}~\bibnamefont {Abbamonte}}, \bibinfo {author} {\bibfnamefont {P.}~\bibnamefont {Pagliuso}}, \emph {et~al.},\ }\bibfield  {title} {\bibinfo {title} {Magnetic polaron formation in {EuZn$_2$P$_2$}},\ }\href@noop {} {\bibfield  {journal} {\bibinfo  {journal} {arXiv preprint arXiv:2504.05494}\ } (\bibinfo {year} {2025})}\BibitemShut {NoStop}%
\bibitem [{\citenamefont {Sichelschmidt}\ \emph {et~al.}(2025)\citenamefont {Sichelschmidt}, \citenamefont {Chailloleau}, \citenamefont {Krebber}, \citenamefont {Mard}, \citenamefont {Krellner},\ and\ \citenamefont {Kliemt}}]{sichelschmidt2025electron}%
  \BibitemOpen
  \bibfield  {author} {\bibinfo {author} {\bibfnamefont {J.}~\bibnamefont {Sichelschmidt}}, \bibinfo {author} {\bibfnamefont {P.}~\bibnamefont {Chailloleau}}, \bibinfo {author} {\bibfnamefont {S.}~\bibnamefont {Krebber}}, \bibinfo {author} {\bibfnamefont {A.~E.}\ \bibnamefont {Mard}}, \bibinfo {author} {\bibfnamefont {C.}~\bibnamefont {Krellner}},\ and\ \bibinfo {author} {\bibfnamefont {K.}~\bibnamefont {Kliemt}},\ }\bibfield  {title} {\bibinfo {title} {Electron spin resonance of eu on triangular layers in {EuT2P2} {(T= Mn, Zn, Cd)}},\ }\href@noop {} {\bibfield  {journal} {\bibinfo  {journal} {arXiv preprint arXiv:2505.06060}\ } (\bibinfo {year} {2025})}\BibitemShut {NoStop}%
\bibitem [{\citenamefont {Nikolic}\ \emph {et~al.}(2011)\citenamefont {Nikolic}, \citenamefont {Vasic}, \citenamefont {Fetahovic}, \citenamefont {Stankovic},\ and\ \citenamefont {Osmokrovic}}]{nikolic2011photodiode}%
  \BibitemOpen
  \bibfield  {author} {\bibinfo {author} {\bibfnamefont {D.}~\bibnamefont {Nikolic}}, \bibinfo {author} {\bibfnamefont {A.}~\bibnamefont {Vasic}}, \bibinfo {author} {\bibfnamefont {I.}~\bibnamefont {Fetahovic}}, \bibinfo {author} {\bibfnamefont {K.}~\bibnamefont {Stankovic}},\ and\ \bibinfo {author} {\bibfnamefont {P.}~\bibnamefont {Osmokrovic}},\ }\bibfield  {title} {\bibinfo {title} {Photodiode behavior in radiation environment},\ }\href@noop {} {\bibfield  {journal} {\bibinfo  {journal} {Scientific Publications of the State University of Novi Pazar Series A}\ }\textbf {\bibinfo {volume} {3}},\ \bibinfo {pages} {27} (\bibinfo {year} {2011})}\BibitemShut {NoStop}%
\bibitem [{\citenamefont {Rosa}\ \emph {et~al.}(2012)\citenamefont {Rosa}, \citenamefont {Adriano}, \citenamefont {Garitezi}, \citenamefont {Ribeiro}, \citenamefont {Fisk},\ and\ \citenamefont {Pagliuso}}]{rosa2012electron}%
  \BibitemOpen
  \bibfield  {author} {\bibinfo {author} {\bibfnamefont {P.}~\bibnamefont {Rosa}}, \bibinfo {author} {\bibfnamefont {C.}~\bibnamefont {Adriano}}, \bibinfo {author} {\bibfnamefont {T.}~\bibnamefont {Garitezi}}, \bibinfo {author} {\bibfnamefont {R.}~\bibnamefont {Ribeiro}}, \bibinfo {author} {\bibfnamefont {Z.}~\bibnamefont {Fisk}},\ and\ \bibinfo {author} {\bibfnamefont {P.}~\bibnamefont {Pagliuso}},\ }\bibfield  {title} {\bibinfo {title} {Electron spin resonance of the intermetallic antiferromagnet {EuIn$_2$As$_2$}},\ }\href@noop {} {\bibfield  {journal} {\bibinfo  {journal} {Physical Review B—Condensed Matter and Materials Physics}\ }\textbf {\bibinfo {volume} {86}},\ \bibinfo {pages} {094408} (\bibinfo {year} {2012})}\BibitemShut {NoStop}%
\bibitem [{\citenamefont {Abragam}\ and\ \citenamefont {Bleaney}(1970)}]{abragam1970electron}%
  \BibitemOpen
  \bibfield  {author} {\bibinfo {author} {\bibfnamefont {A.}~\bibnamefont {Abragam}}\ and\ \bibinfo {author} {\bibfnamefont {B.}~\bibnamefont {Bleaney}},\ }\href {https://books.google.com.br/books?id=SSD7AAAAQBAJ} {\emph {\bibinfo {title} {Electron Paramagnetic Resonance of Transition Ions}}},\ International series of monographs on physics\ (\bibinfo  {publisher} {Clarendon P.},\ \bibinfo {year} {1970})\BibitemShut {NoStop}%
\bibitem [{\citenamefont {Yakovlev}\ and\ \citenamefont {Ossau}(2010)}]{yakovlev2010magnetic}%
  \BibitemOpen
  \bibfield  {author} {\bibinfo {author} {\bibfnamefont {D.~R.}\ \bibnamefont {Yakovlev}}\ and\ \bibinfo {author} {\bibfnamefont {W.}~\bibnamefont {Ossau}},\ }\bibfield  {title} {\bibinfo {title} {Magnetic polarons},\ }in\ \href@noop {} {\emph {\bibinfo {booktitle} {Introduction to the physics of diluted magnetic semiconductors}}}\ (\bibinfo  {publisher} {Springer},\ \bibinfo {year} {2010})\ pp.\ \bibinfo {pages} {221--262}\BibitemShut {NoStop}%
\bibitem [{\citenamefont {Souza}\ \emph {et~al.}(2022)\citenamefont {Souza}, \citenamefont {Thomas}, \citenamefont {Bauer}, \citenamefont {Thompson}, \citenamefont {Ronning}, \citenamefont {Pagliuso},\ and\ \citenamefont {Rosa}}]{souza2022microscopic}%
  \BibitemOpen
  \bibfield  {author} {\bibinfo {author} {\bibfnamefont {J.~C.}\ \bibnamefont {Souza}}, \bibinfo {author} {\bibfnamefont {S.}~\bibnamefont {Thomas}}, \bibinfo {author} {\bibfnamefont {E.}~\bibnamefont {Bauer}}, \bibinfo {author} {\bibfnamefont {J.}~\bibnamefont {Thompson}}, \bibinfo {author} {\bibfnamefont {F.}~\bibnamefont {Ronning}}, \bibinfo {author} {\bibfnamefont {P.}~\bibnamefont {Pagliuso}},\ and\ \bibinfo {author} {\bibfnamefont {P.}~\bibnamefont {Rosa}},\ }\bibfield  {title} {\bibinfo {title} {Microscopic probe of magnetic polarons in antiferromagnetic {Eu$_5$In$_2$Sb$_6$}},\ }\href@noop {} {\bibfield  {journal} {\bibinfo  {journal} {Physical Review B}\ }\textbf {\bibinfo {volume} {105}},\ \bibinfo {pages} {035135} (\bibinfo {year} {2022})}\BibitemShut {NoStop}%
\bibitem [{\citenamefont {Yang}\ \emph {et~al.}(2022)\citenamefont {Yang}, \citenamefont {Zhou},\ and\ \citenamefont {Wang}}]{yang2022topological}%
  \BibitemOpen
  \bibfield  {author} {\bibinfo {author} {\bibfnamefont {M.}~\bibnamefont {Yang}}, \bibinfo {author} {\bibfnamefont {H.}~\bibnamefont {Zhou}},\ and\ \bibinfo {author} {\bibfnamefont {J.}~\bibnamefont {Wang}},\ }\bibfield  {title} {\bibinfo {title} {Topological insulators photodetectors: Preparation, advances and application challenges},\ }\href@noop {} {\bibfield  {journal} {\bibinfo  {journal} {Materials Today Communications}\ }\textbf {\bibinfo {volume} {33}},\ \bibinfo {pages} {104190} (\bibinfo {year} {2022})}\BibitemShut {NoStop}%
\bibitem [{\citenamefont {Enz}\ \emph {et~al.}(1969)\citenamefont {Enz}, \citenamefont {Lems}, \citenamefont {Metselaar}, \citenamefont {Rijnierse},\ and\ \citenamefont {Teale}}]{enz1969photomagnetic}%
  \BibitemOpen
  \bibfield  {author} {\bibinfo {author} {\bibfnamefont {U.}~\bibnamefont {Enz}}, \bibinfo {author} {\bibfnamefont {W.}~\bibnamefont {Lems}}, \bibinfo {author} {\bibfnamefont {R.}~\bibnamefont {Metselaar}}, \bibinfo {author} {\bibfnamefont {P.}~\bibnamefont {Rijnierse}},\ and\ \bibinfo {author} {\bibfnamefont {R.}~\bibnamefont {Teale}},\ }\bibfield  {title} {\bibinfo {title} {Photomagnetic effects},\ }\href@noop {} {\bibfield  {journal} {\bibinfo  {journal} {IEEE Transactions on Magnetics}\ }\textbf {\bibinfo {volume} {5}},\ \bibinfo {pages} {467} (\bibinfo {year} {1969})}\BibitemShut {NoStop}%
\bibitem [{\citenamefont {Garnier}\ \emph {et~al.}(2016)\citenamefont {Garnier}, \citenamefont {Jim{\'e}nez}, \citenamefont {Li}, \citenamefont {Von~Bardeleben}, \citenamefont {Journaux}, \citenamefont {Augenstein}, \citenamefont {Moos}, \citenamefont {Gamer}, \citenamefont {Breher},\ and\ \citenamefont {Lescou{\"e}zec}}]{garnier2016k}%
  \BibitemOpen
  \bibfield  {author} {\bibinfo {author} {\bibfnamefont {D.}~\bibnamefont {Garnier}}, \bibinfo {author} {\bibfnamefont {J.-R.}\ \bibnamefont {Jim{\'e}nez}}, \bibinfo {author} {\bibfnamefont {Y.}~\bibnamefont {Li}}, \bibinfo {author} {\bibfnamefont {J.}~\bibnamefont {Von~Bardeleben}}, \bibinfo {author} {\bibfnamefont {Y.}~\bibnamefont {Journaux}}, \bibinfo {author} {\bibfnamefont {T.}~\bibnamefont {Augenstein}}, \bibinfo {author} {\bibfnamefont {E.}~\bibnamefont {Moos}}, \bibinfo {author} {\bibfnamefont {M.}~\bibnamefont {Gamer}}, \bibinfo {author} {\bibfnamefont {F.}~\bibnamefont {Breher}},\ and\ \bibinfo {author} {\bibfnamefont {R.}~\bibnamefont {Lescou{\"e}zec}},\ }\bibfield  {title} {\bibinfo {title} {${K\subset{[Fe^{II}(Tp)(CN)_3]_4[Co^{III}(^{pz}Tp)]_3[Co^{II}(^{pz}Tp)]}}$: a neutral soluble model complex of photomagnetic prussian blue analogues},\ }\href@noop {} {\bibfield  {journal} {\bibinfo  {journal} {Chemical science}\ }\textbf {\bibinfo {volume} {7}},\ \bibinfo {pages} {4825} (\bibinfo {year}
  {2016})}\BibitemShut {NoStop}%
\bibitem [{\citenamefont {Herrera}\ \emph {et~al.}(2004)\citenamefont {Herrera}, \citenamefont {Marvaud}, \citenamefont {Verdaguer}, \citenamefont {Marrot}, \citenamefont {Kalisz},\ and\ \citenamefont {Mathoniere}}]{herrera2004reversible}%
  \BibitemOpen
  \bibfield  {author} {\bibinfo {author} {\bibfnamefont {J.~M.}\ \bibnamefont {Herrera}}, \bibinfo {author} {\bibfnamefont {V.}~\bibnamefont {Marvaud}}, \bibinfo {author} {\bibfnamefont {M.}~\bibnamefont {Verdaguer}}, \bibinfo {author} {\bibfnamefont {J.}~\bibnamefont {Marrot}}, \bibinfo {author} {\bibfnamefont {M.}~\bibnamefont {Kalisz}},\ and\ \bibinfo {author} {\bibfnamefont {C.}~\bibnamefont {Mathoniere}},\ }\bibfield  {title} {\bibinfo {title} {Reversible photoinduced magnetic properties in the heptanuclear complex {[Mo$^{IV}$(CN)$_2$(CN-CuL)$_6$]$^{8+}$: A Photomagnetic High-Spin Molecule}},\ }\href@noop {} {\bibfield  {journal} {\bibinfo  {journal} {Angewandte Chemie International Edition}\ }\textbf {\bibinfo {volume} {43}},\ \bibinfo {pages} {5468} (\bibinfo {year} {2004})}\BibitemShut {NoStop}%
\bibitem [{\citenamefont {{\'a}O'Connor}\ \emph {et~al.}(1992)\citenamefont {{\'a}O'Connor} \emph {et~al.}}]{ao1992electric}%
  \BibitemOpen
  \bibfield  {author} {\bibinfo {author} {\bibfnamefont {C.~J.}\ \bibnamefont {{\'a}O'Connor}} \emph {et~al.},\ }\bibfield  {title} {\bibinfo {title} {Electric, magnetic, and photomagnetic properties of the amorphous metallic spin-glass {Ni$_3$(SbTe$_3$)$_2$}},\ }\href@noop {} {\bibfield  {journal} {\bibinfo  {journal} {Journal of Materials Chemistry}\ }\textbf {\bibinfo {volume} {2}},\ \bibinfo {pages} {829} (\bibinfo {year} {1992})}\BibitemShut {NoStop}%
\bibitem [{\citenamefont {Wu}\ \emph {et~al.}(1994)\citenamefont {Wu}, \citenamefont {Ren}, \citenamefont {O'Connor}, \citenamefont {Tang}, \citenamefont {Jung}, \citenamefont {Ferr{\'e}},\ and\ \citenamefont {Jamet}}]{wu1994photo}%
  \BibitemOpen
  \bibfield  {author} {\bibinfo {author} {\bibfnamefont {B.}~\bibnamefont {Wu}}, \bibinfo {author} {\bibfnamefont {L.}~\bibnamefont {Ren}}, \bibinfo {author} {\bibfnamefont {C.~J.}\ \bibnamefont {O'Connor}}, \bibinfo {author} {\bibfnamefont {J.}~\bibnamefont {Tang}}, \bibinfo {author} {\bibfnamefont {J.-S.}\ \bibnamefont {Jung}}, \bibinfo {author} {\bibfnamefont {J.}~\bibnamefont {Ferr{\'e}}},\ and\ \bibinfo {author} {\bibfnamefont {J.-P.}\ \bibnamefont {Jamet}},\ }\bibfield  {title} {\bibinfo {title} {Photo-induced magnetic behavior in the amorphous spin-glass material {Co$_3$(SbTe$_3$)$_2$}},\ }\href@noop {} {\bibfield  {journal} {\bibinfo  {journal} {Journal of materials research}\ }\textbf {\bibinfo {volume} {9}},\ \bibinfo {pages} {909} (\bibinfo {year} {1994})}\BibitemShut {NoStop}%
\bibitem [{\citenamefont {O'Connor}\ and\ \citenamefont {Noonan}(1987)}]{o1987photo}%
  \BibitemOpen
  \bibfield  {author} {\bibinfo {author} {\bibfnamefont {C.~J.}\ \bibnamefont {O'Connor}}\ and\ \bibinfo {author} {\bibfnamefont {J.~F.}\ \bibnamefont {Noonan}},\ }\bibfield  {title} {\bibinfo {title} {Photo-induced magnetic bubbles in the spin glass alloy {Fe$_2$SnTe$_4$}},\ }\href@noop {} {\bibfield  {journal} {\bibinfo  {journal} {Journal of Physics and Chemistry of Solids}\ }\textbf {\bibinfo {volume} {48}},\ \bibinfo {pages} {303} (\bibinfo {year} {1987})}\BibitemShut {NoStop}%
\bibitem [{\citenamefont {Chen}\ \emph {et~al.}(2024)\citenamefont {Chen}, \citenamefont {Yang}, \citenamefont {Lu}, \citenamefont {Zhou}, \citenamefont {Ren}, \citenamefont {Cao}, \citenamefont {Dong},\ and\ \citenamefont {Wang}}]{chen2024carrier}%
  \BibitemOpen
  \bibfield  {author} {\bibinfo {author} {\bibfnamefont {X.}~\bibnamefont {Chen}}, \bibinfo {author} {\bibfnamefont {W.}~\bibnamefont {Yang}}, \bibinfo {author} {\bibfnamefont {J.-Y.}\ \bibnamefont {Lu}}, \bibinfo {author} {\bibfnamefont {Z.}~\bibnamefont {Zhou}}, \bibinfo {author} {\bibfnamefont {Z.}~\bibnamefont {Ren}}, \bibinfo {author} {\bibfnamefont {G.-H.}\ \bibnamefont {Cao}}, \bibinfo {author} {\bibfnamefont {S.}~\bibnamefont {Dong}},\ and\ \bibinfo {author} {\bibfnamefont {Z.-C.}\ \bibnamefont {Wang}},\ }\bibfield  {title} {\bibinfo {title} {Carrier-induced transition from antiferromagnetic insulator to ferromagnetic metal in the layered phosphide {EuZn$_2$P$_2$}},\ }\href@noop {} {\bibfield  {journal} {\bibinfo  {journal} {Physical Review B}\ }\textbf {\bibinfo {volume} {109}},\ \bibinfo {pages} {L180410} (\bibinfo {year} {2024})}\BibitemShut {NoStop}%
\bibitem [{\citenamefont {Krebber}\ \emph {et~al.}(2023)\citenamefont {Krebber}, \citenamefont {Kopp}, \citenamefont {Garg}, \citenamefont {Kummer}, \citenamefont {Sichelschmidt}, \citenamefont {Schulz}, \citenamefont {Poelchen}, \citenamefont {Mende}, \citenamefont {Warawa}, \citenamefont {Thomson} \emph {et~al.}}]{krebber2023colossal}%
  \BibitemOpen
  \bibfield  {author} {\bibinfo {author} {\bibfnamefont {S.}~\bibnamefont {Krebber}}, \bibinfo {author} {\bibfnamefont {M.}~\bibnamefont {Kopp}}, \bibinfo {author} {\bibfnamefont {C.}~\bibnamefont {Garg}}, \bibinfo {author} {\bibfnamefont {K.}~\bibnamefont {Kummer}}, \bibinfo {author} {\bibfnamefont {J.}~\bibnamefont {Sichelschmidt}}, \bibinfo {author} {\bibfnamefont {S.}~\bibnamefont {Schulz}}, \bibinfo {author} {\bibfnamefont {G.}~\bibnamefont {Poelchen}}, \bibinfo {author} {\bibfnamefont {M.}~\bibnamefont {Mende}}, \bibinfo {author} {\bibfnamefont {K.}~\bibnamefont {Warawa}}, \bibinfo {author} {\bibfnamefont {M.~D.}\ \bibnamefont {Thomson}}, \emph {et~al.},\ }\bibfield  {title} {\bibinfo {title} {Colossal magnetoresistance in {EuZn$_2$P$_2$} and its electronic and magnetic structure},\ }\href@noop {} {\bibfield  {journal} {\bibinfo  {journal} {arXiv preprint arXiv:2302.14539}\ } (\bibinfo {year} {2023})}\BibitemShut {NoStop}%
\bibitem [{\citenamefont {Yang}\ \emph {et~al.}(2004)\citenamefont {Yang}, \citenamefont {Bao}, \citenamefont {Tan},\ and\ \citenamefont {Zhang}}]{yang2004magnetic}%
  \BibitemOpen
  \bibfield  {author} {\bibinfo {author} {\bibfnamefont {Z.}~\bibnamefont {Yang}}, \bibinfo {author} {\bibfnamefont {X.}~\bibnamefont {Bao}}, \bibinfo {author} {\bibfnamefont {S.}~\bibnamefont {Tan}},\ and\ \bibinfo {author} {\bibfnamefont {Y.}~\bibnamefont {Zhang}},\ }\bibfield  {title} {\bibinfo {title} {Magnetic polaron conduction in the colossal magnetoresistance material {Fe$_{1-x}$Cd$_x$Cr$_2$S$_4$}},\ }\href@noop {} {\bibfield  {journal} {\bibinfo  {journal} {Physical Review B}\ }\textbf {\bibinfo {volume} {69}},\ \bibinfo {pages} {144407} (\bibinfo {year} {2004})}\BibitemShut {NoStop}%
\bibitem [{\citenamefont {Ying}\ \emph {et~al.}(2018)\citenamefont {Ying}, \citenamefont {Tang}, \citenamefont {Chen}, \citenamefont {Chen},\ and\ \citenamefont {Struzhkin}}]{ying2018coexistence}%
  \BibitemOpen
  \bibfield  {author} {\bibinfo {author} {\bibfnamefont {J.}~\bibnamefont {Ying}}, \bibinfo {author} {\bibfnamefont {L.}~\bibnamefont {Tang}}, \bibinfo {author} {\bibfnamefont {F.}~\bibnamefont {Chen}}, \bibinfo {author} {\bibfnamefont {X.}~\bibnamefont {Chen}},\ and\ \bibinfo {author} {\bibfnamefont {V.~V.}\ \bibnamefont {Struzhkin}},\ }\bibfield  {title} {\bibinfo {title} {Coexistence of metallic and insulating channels in compressed {YbB$_6$}},\ }\href@noop {} {\bibfield  {journal} {\bibinfo  {journal} {Physical Review B}\ }\textbf {\bibinfo {volume} {97}},\ \bibinfo {pages} {121101} (\bibinfo {year} {2018})}\BibitemShut {NoStop}%
\bibitem [{\citenamefont {Li}\ \emph {et~al.}(2020)\citenamefont {Li}, \citenamefont {Sun}, \citenamefont {Kurdak},\ and\ \citenamefont {Allen}}]{li2020emergent}%
  \BibitemOpen
  \bibfield  {author} {\bibinfo {author} {\bibfnamefont {L.}~\bibnamefont {Li}}, \bibinfo {author} {\bibfnamefont {K.}~\bibnamefont {Sun}}, \bibinfo {author} {\bibfnamefont {C.}~\bibnamefont {Kurdak}},\ and\ \bibinfo {author} {\bibfnamefont {J.}~\bibnamefont {Allen}},\ }\bibfield  {title} {\bibinfo {title} {Emergent mystery in the kondo insulator samarium hexaboride},\ }\href@noop {} {\bibfield  {journal} {\bibinfo  {journal} {Nature Reviews Physics}\ }\textbf {\bibinfo {volume} {2}},\ \bibinfo {pages} {463} (\bibinfo {year} {2020})}\BibitemShut {NoStop}%
\bibitem [{\citenamefont {Usachov}\ \emph {et~al.}(2024)\citenamefont {Usachov}, \citenamefont {Krebber}, \citenamefont {Bokai}, \citenamefont {Tarasov}, \citenamefont {Kopp}, \citenamefont {Garg}, \citenamefont {Virovets}, \citenamefont {M{\"u}ller}, \citenamefont {Mende}, \citenamefont {Poelchen} \emph {et~al.}}]{usachov2024magnetism}%
  \BibitemOpen
  \bibfield  {author} {\bibinfo {author} {\bibfnamefont {D.~Y.}\ \bibnamefont {Usachov}}, \bibinfo {author} {\bibfnamefont {S.}~\bibnamefont {Krebber}}, \bibinfo {author} {\bibfnamefont {K.~A.}\ \bibnamefont {Bokai}}, \bibinfo {author} {\bibfnamefont {A.~V.}\ \bibnamefont {Tarasov}}, \bibinfo {author} {\bibfnamefont {M.}~\bibnamefont {Kopp}}, \bibinfo {author} {\bibfnamefont {C.}~\bibnamefont {Garg}}, \bibinfo {author} {\bibfnamefont {A.}~\bibnamefont {Virovets}}, \bibinfo {author} {\bibfnamefont {J.}~\bibnamefont {M{\"u}ller}}, \bibinfo {author} {\bibfnamefont {M.}~\bibnamefont {Mende}}, \bibinfo {author} {\bibfnamefont {G.}~\bibnamefont {Poelchen}}, \emph {et~al.},\ }\bibfield  {title} {\bibinfo {title} {Magnetism, heat capacity, and electronic structure of {EuCd$_2$P$_2$} in view of its colossal magnetoresistance},\ }\href@noop {} {\bibfield  {journal} {\bibinfo  {journal} {Physical Review B}\ }\textbf {\bibinfo {volume} {109}},\ \bibinfo {pages} {104421} (\bibinfo {year} {2024})}\BibitemShut {NoStop}%
\bibitem [{\citenamefont {Rosa}\ \emph {et~al.}(2020)\citenamefont {Rosa}, \citenamefont {Xu}, \citenamefont {Rahn}, \citenamefont {Souza}, \citenamefont {Kushwaha}, \citenamefont {Veiga}, \citenamefont {Bombardi}, \citenamefont {Thomas}, \citenamefont {Janoschek}, \citenamefont {Bauer} \emph {et~al.}}]{rosa2020colossal}%
  \BibitemOpen
  \bibfield  {author} {\bibinfo {author} {\bibfnamefont {P.}~\bibnamefont {Rosa}}, \bibinfo {author} {\bibfnamefont {Y.}~\bibnamefont {Xu}}, \bibinfo {author} {\bibfnamefont {M.}~\bibnamefont {Rahn}}, \bibinfo {author} {\bibfnamefont {J.}~\bibnamefont {Souza}}, \bibinfo {author} {\bibfnamefont {S.}~\bibnamefont {Kushwaha}}, \bibinfo {author} {\bibfnamefont {L.}~\bibnamefont {Veiga}}, \bibinfo {author} {\bibfnamefont {A.}~\bibnamefont {Bombardi}}, \bibinfo {author} {\bibfnamefont {S.}~\bibnamefont {Thomas}}, \bibinfo {author} {\bibfnamefont {M.}~\bibnamefont {Janoschek}}, \bibinfo {author} {\bibfnamefont {E.}~\bibnamefont {Bauer}}, \emph {et~al.},\ }\bibfield  {title} {\bibinfo {title} {Colossal magnetoresistance in a nonsymmorphic antiferromagnetic insulator},\ }\href@noop {} {\bibfield  {journal} {\bibinfo  {journal} {npj Quantum Materials}\ }\textbf {\bibinfo {volume} {5}},\ \bibinfo {pages} {52} (\bibinfo {year} {2020})}\BibitemShut {NoStop}%
\bibitem [{\citenamefont {Feher}\ and\ \citenamefont {Kip}(1955)}]{feher1955electron}%
  \BibitemOpen
  \bibfield  {author} {\bibinfo {author} {\bibfnamefont {G.}~\bibnamefont {Feher}}\ and\ \bibinfo {author} {\bibfnamefont {A.}~\bibnamefont {Kip}},\ }\bibfield  {title} {\bibinfo {title} {Electron spin resonance absorption in metals. i. experimental},\ }\href@noop {} {\bibfield  {journal} {\bibinfo  {journal} {Physical Review}\ }\textbf {\bibinfo {volume} {98}},\ \bibinfo {pages} {337} (\bibinfo {year} {1955})}\BibitemShut {NoStop}%
\bibitem [{\citenamefont {Kao}(2004)}]{kao2004dielectric}%
  \BibitemOpen
  \bibfield  {author} {\bibinfo {author} {\bibfnamefont {K.~C.}\ \bibnamefont {Kao}},\ }\href@noop {} {\emph {\bibinfo {title} {Dielectric phenomena in solids}}}\ (\bibinfo  {publisher} {Elsevier},\ \bibinfo {year} {2004})\BibitemShut {NoStop}%
\bibitem [{\citenamefont {Goldman}\ \emph {et~al.}(1978)\citenamefont {Goldman}, \citenamefont {Kalikstein},\ and\ \citenamefont {Kramer}}]{goldman1978dember}%
  \BibitemOpen
  \bibfield  {author} {\bibinfo {author} {\bibfnamefont {S.~R.}\ \bibnamefont {Goldman}}, \bibinfo {author} {\bibfnamefont {K.}~\bibnamefont {Kalikstein}},\ and\ \bibinfo {author} {\bibfnamefont {B.}~\bibnamefont {Kramer}},\ }\bibfield  {title} {\bibinfo {title} {Dember-effect theory},\ }\href@noop {} {\bibfield  {journal} {\bibinfo  {journal} {Journal of Applied Physics}\ }\textbf {\bibinfo {volume} {49}},\ \bibinfo {pages} {2849} (\bibinfo {year} {1978})}\BibitemShut {NoStop}%
\bibitem [{\citenamefont {Kugel}\ \emph {et~al.}(2008)\citenamefont {Kugel}, \citenamefont {Rakhmanov}, \citenamefont {Sboychakov}, \citenamefont {Kagan},\ and\ \citenamefont {Ogarkov}}]{kugel2008structure}%
  \BibitemOpen
  \bibfield  {author} {\bibinfo {author} {\bibfnamefont {K.}~\bibnamefont {Kugel}}, \bibinfo {author} {\bibfnamefont {A.}~\bibnamefont {Rakhmanov}}, \bibinfo {author} {\bibfnamefont {A.}~\bibnamefont {Sboychakov}}, \bibinfo {author} {\bibfnamefont {M.~Y.}\ \bibnamefont {Kagan}},\ and\ \bibinfo {author} {\bibfnamefont {S.}~\bibnamefont {Ogarkov}},\ }\bibfield  {title} {\bibinfo {title} {The structure of magnetic polarons in doped antiferromagnetic insulators},\ }\href@noop {} {\bibfield  {journal} {\bibinfo  {journal} {Physica B: Condensed Matter}\ }\textbf {\bibinfo {volume} {403}},\ \bibinfo {pages} {1353} (\bibinfo {year} {2008})}\BibitemShut {NoStop}%
\bibitem [{\citenamefont {Abragam}\ and\ \citenamefont {Bleaney}(2012)}]{abragam2012electron}%
  \BibitemOpen
  \bibfield  {author} {\bibinfo {author} {\bibfnamefont {A.}~\bibnamefont {Abragam}}\ and\ \bibinfo {author} {\bibfnamefont {B.}~\bibnamefont {Bleaney}},\ }\href@noop {} {\emph {\bibinfo {title} {Electron paramagnetic resonance of transition ions}}}\ (\bibinfo  {publisher} {OUP Oxford},\ \bibinfo {year} {2012})\BibitemShut {NoStop}%
\bibitem [{\citenamefont {Ghosh}\ \emph {et~al.}(2022)\citenamefont {Ghosh}, \citenamefont {Lane}, \citenamefont {Ronning}, \citenamefont {Bauer}, \citenamefont {Thompson}, \citenamefont {Zhu}, \citenamefont {Rosa},\ and\ \citenamefont {Thomas}}]{ghosh2022colossal}%
  \BibitemOpen
  \bibfield  {author} {\bibinfo {author} {\bibfnamefont {S.}~\bibnamefont {Ghosh}}, \bibinfo {author} {\bibfnamefont {C.}~\bibnamefont {Lane}}, \bibinfo {author} {\bibfnamefont {F.}~\bibnamefont {Ronning}}, \bibinfo {author} {\bibfnamefont {E.~D.}\ \bibnamefont {Bauer}}, \bibinfo {author} {\bibfnamefont {J.~D.}\ \bibnamefont {Thompson}}, \bibinfo {author} {\bibfnamefont {J.-X.}\ \bibnamefont {Zhu}}, \bibinfo {author} {\bibfnamefont {P.}~\bibnamefont {Rosa}},\ and\ \bibinfo {author} {\bibfnamefont {S.~M.}\ \bibnamefont {Thomas}},\ }\bibfield  {title} {\bibinfo {title} {Colossal piezoresistance in narrow-gap {Eu$_5$In$_2$Sb$_6$}},\ }\href@noop {} {\bibfield  {journal} {\bibinfo  {journal} {Physical Review B}\ }\textbf {\bibinfo {volume} {106}},\ \bibinfo {pages} {045110} (\bibinfo {year} {2022})}\BibitemShut {NoStop}%
\bibitem [{\citenamefont {Majumdar}\ and\ \citenamefont {Littlewood}(1998)}]{majumdar1998magnetoresistance}%
  \BibitemOpen
  \bibfield  {author} {\bibinfo {author} {\bibfnamefont {P.}~\bibnamefont {Majumdar}}\ and\ \bibinfo {author} {\bibfnamefont {P.}~\bibnamefont {Littlewood}},\ }\bibfield  {title} {\bibinfo {title} {Magnetoresistance in mn pyrochlore: electrical transport in a low carrier density ferromagnet},\ }\href@noop {} {\bibfield  {journal} {\bibinfo  {journal} {Physical review letters}\ }\textbf {\bibinfo {volume} {81}},\ \bibinfo {pages} {1314} (\bibinfo {year} {1998})}\BibitemShut {NoStop}%
\bibitem [{\citenamefont {Xu}\ \emph {et~al.}(2019)\citenamefont {Xu}, \citenamefont {Song}, \citenamefont {Wang}, \citenamefont {Weng},\ and\ \citenamefont {Dai}}]{xu2019higher}%
  \BibitemOpen
  \bibfield  {author} {\bibinfo {author} {\bibfnamefont {Y.}~\bibnamefont {Xu}}, \bibinfo {author} {\bibfnamefont {Z.}~\bibnamefont {Song}}, \bibinfo {author} {\bibfnamefont {Z.}~\bibnamefont {Wang}}, \bibinfo {author} {\bibfnamefont {H.}~\bibnamefont {Weng}},\ and\ \bibinfo {author} {\bibfnamefont {X.}~\bibnamefont {Dai}},\ }\bibfield  {title} {\bibinfo {title} {Higher-order topology of the axion insulator {EuIn$_2$As$_2$}},\ }\href@noop {} {\bibfield  {journal} {\bibinfo  {journal} {Physical review letters}\ }\textbf {\bibinfo {volume} {122}},\ \bibinfo {pages} {256402} (\bibinfo {year} {2019})}\BibitemShut {NoStop}%
\bibitem [{\citenamefont {Singh}\ \emph {et~al.}(2023)\citenamefont {Singh}, \citenamefont {Dan}, \citenamefont {Ptok}, \citenamefont {Zaleski}, \citenamefont {Pavlosiuk}, \citenamefont {Wi{\'s}niewski},\ and\ \citenamefont {Kaczorowski}}]{singh2023superexchange}%
  \BibitemOpen
  \bibfield  {author} {\bibinfo {author} {\bibfnamefont {K.}~\bibnamefont {Singh}}, \bibinfo {author} {\bibfnamefont {S.}~\bibnamefont {Dan}}, \bibinfo {author} {\bibfnamefont {A.}~\bibnamefont {Ptok}}, \bibinfo {author} {\bibfnamefont {T.}~\bibnamefont {Zaleski}}, \bibinfo {author} {\bibfnamefont {O.}~\bibnamefont {Pavlosiuk}}, \bibinfo {author} {\bibfnamefont {P.}~\bibnamefont {Wi{\'s}niewski}},\ and\ \bibinfo {author} {\bibfnamefont {D.}~\bibnamefont {Kaczorowski}},\ }\bibfield  {title} {\bibinfo {title} {Superexchange interaction in insulating {EuZn$_2$P$_2$}},\ }\href@noop {} {\bibfield  {journal} {\bibinfo  {journal} {Physical Review B}\ }\textbf {\bibinfo {volume} {108}},\ \bibinfo {pages} {054402} (\bibinfo {year} {2023})}\BibitemShut {NoStop}%
\bibitem [{\citenamefont {Pierantozzi}\ \emph {et~al.}(2022)\citenamefont {Pierantozzi}, \citenamefont {De~Vita}, \citenamefont {Bigi}, \citenamefont {Gui}, \citenamefont {Tien}, \citenamefont {Mondal}, \citenamefont {Mazzola}, \citenamefont {Fujii}, \citenamefont {Vobornik}, \citenamefont {Vinai} \emph {et~al.}}]{pierantozzi2022evidence}%
  \BibitemOpen
  \bibfield  {author} {\bibinfo {author} {\bibfnamefont {G.~M.}\ \bibnamefont {Pierantozzi}}, \bibinfo {author} {\bibfnamefont {A.}~\bibnamefont {De~Vita}}, \bibinfo {author} {\bibfnamefont {C.}~\bibnamefont {Bigi}}, \bibinfo {author} {\bibfnamefont {X.}~\bibnamefont {Gui}}, \bibinfo {author} {\bibfnamefont {H.-J.}\ \bibnamefont {Tien}}, \bibinfo {author} {\bibfnamefont {D.}~\bibnamefont {Mondal}}, \bibinfo {author} {\bibfnamefont {F.}~\bibnamefont {Mazzola}}, \bibinfo {author} {\bibfnamefont {J.}~\bibnamefont {Fujii}}, \bibinfo {author} {\bibfnamefont {I.}~\bibnamefont {Vobornik}}, \bibinfo {author} {\bibfnamefont {G.}~\bibnamefont {Vinai}}, \emph {et~al.},\ }\bibfield  {title} {\bibinfo {title} {Evidence of magnetism-induced topological protection in the axion insulator candidate {EuSn$_2$P$_2$}},\ }\href@noop {} {\bibfield  {journal} {\bibinfo  {journal} {Proceedings of the National Academy of Sciences}\ }\textbf {\bibinfo {volume} {119}},\ \bibinfo {pages} {e2116575119} (\bibinfo {year}
  {2022})}\BibitemShut {NoStop}%
\bibitem [{\citenamefont {Riberolles}\ \emph {et~al.}(2021)\citenamefont {Riberolles}, \citenamefont {Trevisan}, \citenamefont {Kuthanazhi}, \citenamefont {Heitmann}, \citenamefont {Ye}, \citenamefont {Johnston}, \citenamefont {Bud’ko}, \citenamefont {Ryan}, \citenamefont {Canfield}, \citenamefont {Kreyssig} \emph {et~al.}}]{riberolles2021magnetic}%
  \BibitemOpen
  \bibfield  {author} {\bibinfo {author} {\bibfnamefont {S.~X.}\ \bibnamefont {Riberolles}}, \bibinfo {author} {\bibfnamefont {T.~V.}\ \bibnamefont {Trevisan}}, \bibinfo {author} {\bibfnamefont {B.}~\bibnamefont {Kuthanazhi}}, \bibinfo {author} {\bibfnamefont {T.}~\bibnamefont {Heitmann}}, \bibinfo {author} {\bibfnamefont {F.}~\bibnamefont {Ye}}, \bibinfo {author} {\bibfnamefont {D.}~\bibnamefont {Johnston}}, \bibinfo {author} {\bibfnamefont {S.}~\bibnamefont {Bud’ko}}, \bibinfo {author} {\bibfnamefont {D.}~\bibnamefont {Ryan}}, \bibinfo {author} {\bibfnamefont {P.}~\bibnamefont {Canfield}}, \bibinfo {author} {\bibfnamefont {A.}~\bibnamefont {Kreyssig}}, \emph {et~al.},\ }\bibfield  {title} {\bibinfo {title} {Magnetic crystalline-symmetry-protected axion electrodynamics and field-tunable unpinned dirac cones in {EuIn$_2$As$_2$}},\ }\href@noop {} {\bibfield  {journal} {\bibinfo  {journal} {Nature communications}\ }\textbf {\bibinfo {volume} {12}},\ \bibinfo {pages} {999} (\bibinfo {year} {2021})}\BibitemShut
  {NoStop}%
\bibitem [{\citenamefont {Hennion}\ \emph {et~al.}(2000)\citenamefont {Hennion}, \citenamefont {Moussa}, \citenamefont {Biotteau}, \citenamefont {Rodriguez-Carvajal}, \citenamefont {Pinsard},\ and\ \citenamefont {Revcolevschi}}]{hennion2000evidence}%
  \BibitemOpen
  \bibfield  {author} {\bibinfo {author} {\bibfnamefont {M.}~\bibnamefont {Hennion}}, \bibinfo {author} {\bibfnamefont {F.}~\bibnamefont {Moussa}}, \bibinfo {author} {\bibfnamefont {G.}~\bibnamefont {Biotteau}}, \bibinfo {author} {\bibfnamefont {J.}~\bibnamefont {Rodriguez-Carvajal}}, \bibinfo {author} {\bibfnamefont {L.}~\bibnamefont {Pinsard}},\ and\ \bibinfo {author} {\bibfnamefont {A.}~\bibnamefont {Revcolevschi}},\ }\bibfield  {title} {\bibinfo {title} {Evidence of anisotropic magnetic polarons in {La$_{0.94}$Sr$_{0.06}$MnO$_3$} by neutron scattering and comparison with {Ca}-doped manganites},\ }\href@noop {} {\bibfield  {journal} {\bibinfo  {journal} {Physical Review B}\ }\textbf {\bibinfo {volume} {61}},\ \bibinfo {pages} {9513} (\bibinfo {year} {2000})}\BibitemShut {NoStop}%
\bibitem [{\citenamefont {de~Jongh}(2012)}]{de2012magnetic}%
  \BibitemOpen
  \bibfield  {author} {\bibinfo {author} {\bibfnamefont {L.~J.}\ \bibnamefont {de~Jongh}},\ }\href@noop {} {\emph {\bibinfo {title} {Magnetic properties of layered transition metal compounds}}},\ Vol.~\bibinfo {volume} {9}\ (\bibinfo  {publisher} {Springer Science \& Business Media},\ \bibinfo {year} {2012})\BibitemShut {NoStop}%
\bibitem [{\citenamefont {Richards}\ and\ \citenamefont {Salamon}(1974)}]{richards1974exchange}%
  \BibitemOpen
  \bibfield  {author} {\bibinfo {author} {\bibfnamefont {P.~M.}\ \bibnamefont {Richards}}\ and\ \bibinfo {author} {\bibfnamefont {M.}~\bibnamefont {Salamon}},\ }\bibfield  {title} {\bibinfo {title} {Exchange narrowing of electron spin resonance in a two-dimensional system},\ }\href@noop {} {\bibfield  {journal} {\bibinfo  {journal} {Physical Review B}\ }\textbf {\bibinfo {volume} {9}},\ \bibinfo {pages} {32} (\bibinfo {year} {1974})}\BibitemShut {NoStop}%
\bibitem [{\citenamefont {Xin}\ \emph {et~al.}(2018)\citenamefont {Xin}, \citenamefont {Li}, \citenamefont {He}, \citenamefont {Su}, \citenamefont {Jiang}, \citenamefont {Huang}, \citenamefont {Zhou}, \citenamefont {Liu},\ and\ \citenamefont {Tian}}]{xin2018black}%
  \BibitemOpen
  \bibfield  {author} {\bibinfo {author} {\bibfnamefont {W.}~\bibnamefont {Xin}}, \bibinfo {author} {\bibfnamefont {X.-K.}\ \bibnamefont {Li}}, \bibinfo {author} {\bibfnamefont {X.-L.}\ \bibnamefont {He}}, \bibinfo {author} {\bibfnamefont {B.-W.}\ \bibnamefont {Su}}, \bibinfo {author} {\bibfnamefont {X.-Q.}\ \bibnamefont {Jiang}}, \bibinfo {author} {\bibfnamefont {K.-X.}\ \bibnamefont {Huang}}, \bibinfo {author} {\bibfnamefont {X.-F.}\ \bibnamefont {Zhou}}, \bibinfo {author} {\bibfnamefont {Z.-B.}\ \bibnamefont {Liu}},\ and\ \bibinfo {author} {\bibfnamefont {J.-G.}\ \bibnamefont {Tian}},\ }\bibfield  {title} {\bibinfo {title} {Black-phosphorus-based orientation-induced diodes},\ }\href@noop {} {\bibfield  {journal} {\bibinfo  {journal} {Advanced Materials}\ }\textbf {\bibinfo {volume} {30}},\ \bibinfo {pages} {1704653} (\bibinfo {year} {2018})}\BibitemShut {NoStop}%
\end{thebibliography}%

\end{document}